\documentclass[12pt]{article}

\input epsf
\usepackage{amssymb,amsmath,wrapfig}
\usepackage[matrix,arrow,curve]{xy}
\usepackage{cite}
\usepackage{pdfsync}
\usepackage{graphicx,color}
\usepackage[debug,pageanchor=false]{hyperref}
\definecolor{link}{rgb}{.8,.15,.1}
\hypersetup{colorlinks=true,linkcolor=link,citecolor=link,urlcolor=link,linktocpage}

\makeatletter
\@addtoreset{equation}{section}
\makeatother

\def\rr {{\Bbb R}}
\def\cc {{\Bbb C}}
\def\pp {{\Bbb P}}

\def\del {\partial}

\def\del {\partial}

\def\sla#1{\rlap{\begin{picture}(10,10)
\put(0,0){\line(1,1){10}}
\end{picture} }#1}
\def\slash#1#2#3{\rlap{\begin{picture}(10,10)
\put(0,0){\line(#1,1){#2}}
\end{picture}}#3}

\begin{document}

	       \begin{titlepage}
	       \begin{center}

	       \vskip .3in \noindent

	       {\Large \bf{Generalized structures of ten-dimensional supersymmetric solutions}}

	       \bigskip

		 Alessandro Tomasiello\\

	       \bigskip
			 Dipartimento di Fisica, Universit\`a di Milano-Bicocca, I-20126 Milano, Italy\\
	       and\\
	       INFN, sezione di Milano-Bicocca,
	       I-20126 Milano, Italy

	       \vskip .5in
	       {\bf Abstract }
	       \vskip .1in

	       \end{center}

	\hyphenation{dif-fer-en-tial}
	
	Four-dimensional supersymmetric type II string theory vacua can be described elegantly in terms of pure spinors on the generalized tangent bundle $T \oplus T^*$. In this paper, we apply the same techniques to any ten-dimensional supersymmetric solution (not necessarily involving a factor with an ${\rm AdS}_4$ or Minkowski$_4$ metric) in type II theories. We find a system of differential equations in terms of a form describing a ``generalized ISpin(7) structure''. This system is equivalent to unbroken supersymmetry, in both IIA and IIB. One of the equations reproduces in one fell swoop all the pure spinors equations for four-dimensional vacua. 		
	       \vfill
	       \eject

	       \end{titlepage}	

\hypersetup{pageanchor=true}

       \tableofcontents

\section{Introduction} 
\label{sec:intro}

Differential forms are in many respects easier to deal with than symmetric tensors. Gravity is usually described in terms of a symmetric tensor $g_{MN}$. It is an old idea that it might be understood better  if written in terms of forms. Initially this was done in the hope that it might help with quantization; in four dimensions, one can express ordinary general relativity in terms of self-dual two-forms \cite{israel,plebanski,ashtekar,capovilla-dell-jacobson-mason}. More recently, it proved useful in finding classical solutions to various supergravity theories. In this case, the forms are additional data defined by the fermionic supersymmetry parameters of the supergravity theory. Mathematically, they define a reduction of the structure group of the tangent bundle $T$ to a certain group $G$, which is nothing but their little group (or stabilizer). These so-called $G$-structures have been used to reformulate the supersymmetry conditions more efficiently, starting from \cite{gauntlett-pakis,gauntlett-martelli-pakis-waldram}.

In eleven-dimensional supergravity, there is only one supersymmetry parameter $\epsilon$, which can define two possible structure groups \cite{gauntlett-pakis,gauntlett-gutowski-pakis}.  In type II theories, each of the two supersymmetry parameters $\epsilon_{1,2}$ has its own stabilizer, and these can intersect in various ways. This gives rise to a variety of $G$-structures \cite{hackettjones-smith,saffin,gran-gutowski-papadopoulos,gran-gutowski-papadopoulos-2}. Moreover, for a supersymmetric solution the stabilizer of $\epsilon_{1,2}$ need not be the same everywhere: it can jump to a higher group on some locus of spacetime. Because of all this, a complete classification quickly becomes complicated.

A possible reaction to this is to work on $T\oplus T^*$, the direct sum of the tangent and cotangent bundles. This approach has been useful for four-dimensional ``vacua'' (namely, solutions of the form Minkowski$_4\times M_6$ or AdS$_4\times M_6$, with $M_6$ an arbitrary manifold). Here, the stabilizer in $T$ of the two spinors $\epsilon_{1,2}$ can be SU(2), SU(3), or it can be generically SU(2) and jump to SU(3) on some loci. On $T \oplus T^*$, however, the stabilizer is always ${\rm SU}(3) \times {\rm SU}(3)$. This ``generalized structure'' allows then a unified treatment. It can be described in an alternative way by two differential forms $\phi_\pm$, sometimes called ``pure spinors''. The conditions for unbroken supersymmetry can then be summarized elegantly in terms of the $\phi_\pm$ \cite{gmpt2,gmpt3}; the system consists of three equations (see (\ref{eq:psp46})). Interestingly, of these, (\ref{eq:psp46closed}) had been studied by mathematicians before the physical application became clear \cite{hitchin-gcy,gualtieri}.

This success suggests one should be able to apply similar techniques to other types of supersymmetric solutions, beyond four-dimensional vacua. There are several instances where this has been attempted. For example, one can look at other dimensions: solutions of the form Minkowski$_d\times M_{10-d}$ or AdS$_d\times M_{10-d}$. A complete reformulation was achieved for $d=3$ in \cite{haack-lust-martucci-t,smyth-vaula}, and for $d=6$ in \cite{lust-patalong-tsimpis}; a set of necessary conditions was found for $d=1$ in \cite{koerber-martucci-ads} and for all even $d$ in \cite{lust-patalong-tsimpis}. Or, staying in four dimension, one can look for solutions where the spacetime geometry is no longer Minkowski$_4$ or AdS$_4$. For example, a particularly natural geometry to consider is that of a spherically symmetric black hole. This was considered in \cite{hmt}, with some Ansatz along the way. 

There are of course other classes of interesting supersymmetric solutions. It would be interesting, for example, to have a classification of supersymmetric Lifschitz solutions (which are interesting as holographic duals to scale-invariant non-relativistic field theories~\cite{kachru-liu-mulligan}), or to generalize the results of \cite{hmt} to multi-center or asymptotically AdS black holes.

At present, however, every time one is interested in a new class of solutions one has to start from scratch; one first needs to derive some differential equations on the relevant differential forms, then --- much more painfully --- one needs to prove that these equations are equivalent to the conditions on $\epsilon_{1,2}$ for preserved supersymmetry. It would be nice to have a result which applies to \emph{any} type of supersymmetric solution; this would combine the advantages of the $G$-structure papers \cite{gauntlett-pakis,gauntlett-gutowski-pakis,hackettjones-smith,saffin,gran-gutowski-papadopoulos,gran-gutowski-papadopoulos-2}, which make no Ansatz on the metric, with the advantages of the generalized geometry approach \cite{gmpt2,gmpt3}, which unifies the various possibilities for the stabilizers.  

In this paper, we find such a result. For type II supergravity, we give a system of differential equations ((\ref{eq:susy10}) below) which is equivalent to supersymmetry, without any Ansatz on the ten-dimensional metric or on any of the other field. The system is essentially identical in IIA and IIB. 

The geometrical data appearing in the system are a single differential form $\Phi$ (even in IIA, odd in IIB, but otherwise of mixed degree) and two sections $e_{+_1}\cdot, \cdot e_{+_2}$ of $T \oplus T^*$ (our notation will be explained in section \ref{sec:gen}). The form $\Phi$ is not a pure spinor. For four-dimensional vacua, however, it does reduce to a certain sum of the pure spinors $\phi_\pm$ mentioned earlier (see (\ref{eq:Phi46}) below). On $T \oplus T^*$, $\Phi$ defines a complicated structure group, (\ref{eq:genISpin}); this contains two copies of the ``inhomogeneous Spin(7)'' group, ${\rm ISpin}(7)\equiv {\rm Spin}(7)\ltimes \rr^8$, which is the structure group defined by a spinor in ten dimensions (see for example \cite{figueroaofarrill}). Since the group in (\ref{eq:genISpin}) is not a subgroup of ${\rm Spin}(9,1)\times {\rm Spin}(9,1)$, however, $\Phi$ is not enough by itself to determine a metric, as we show in section \ref{sec:gen}. This is why it has to be supplemented by two sections of $T \oplus T^*$. 

Among the differential equations, (\ref{eq:psp10}) is particularly nice. When specialized to four-dimensional vacuum solutions, it is easy to see that it implies all the pure spinor equations of \cite{gmpt2} in one go. This equation is very similar to \cite[Eq.~(A.27)]{koerber-martucci-ads}; perhaps not surprisingly, since their setup ($\rr\times M_9$) is already very general. Deriving (\ref{eq:psp10}) from supersymmetry is in fact even easier (see appendix \ref{sub:psp10}) than deriving the pure spinor equations for four-dimensional vacua (see \cite[App.~A]{gmpt3}). A far harder task, however, is establishing whether it is equivalent to, and not only implied by, supersymmetry. It is not; this is why we had to supplement (\ref{eq:psp10}) by (\ref{eq:LK}), (\ref{eq:++1}), (\ref{eq:++2}). The latter two equations were particularly hard to find, basically because the two sections $e_{+_1}\cdot$ and $\cdot e_{+_2}$ of $T \oplus T^*$ are not defined directly by the spinors $\epsilon_i$. It would be interesting to find alternative sets of equations complementary to (\ref{eq:psp10}). The formalism set up in this paper will hopefully be helpful in doing that.\footnote{Some useful hints might also come from efforts towards reformulating type II supergravity using generalized geometry; for a recent example, see \cite{coimbra-stricklandconstable-waldram}.}

In section \ref{sec:gen} we will look at the structure groups defined on $T$ and $T \oplus T^*$ by the two supersymmetry parameters $\epsilon_{1,2}$, and isolate the geometrical objects $(\Phi,e_{+_1}\cdot, \cdot e_{+_2})$ that will summarize for us the data of the metric, of the $B$ field and of the $\epsilon_{1,2}$. This long section is summarized in section \ref{sub:sumgen}. In section \ref{sec:diff}, we will describe the system (\ref{eq:susy10}) of differential equations which reformulates the requirement of unbroken supersymmetry in terms of the $(\Phi,e_{+_1}\cdot, \cdot e_{+_2})$; its derivation is hidden in section \ref{app:suff}. In section \ref{sec:examples} we will show how the system (\ref{eq:susy10}) reproduces earlier results about four- and three-dimensional Minkowski vacuum solutions. In both cases, equation (\ref{eq:psp10}) reproduces all the ``pure spinor'' equations. The additional equations (\ref{eq:++1}), (\ref{eq:++2}) are redundant for four-dimensional vacua, while for three-dimensional vacua they reproduce a peculiar algebraic constraint that was found in \cite{haack-lust-martucci-t,smyth-vaula}.

	
\section{Generalized ten-dimensional structures} 
\label{sec:gen}

In this section, we will describe how to encode the data of the metric, of the $B$ field and of the supersymmetry parameters in a set of differential forms. A summary of these results can be found in section \ref{sub:sumgen}. These forms are the ones on which we will impose differential equations in section \ref{sec:diff}.

\subsection{Geometry defined by one spinor} 
\label{sub:eps}

Recall that the parameters for the supersymmetry transformations of type II supergravity are two ten-dimensional Majorana--Weyl spinors $\epsilon_1$, $\epsilon_2$. In type IIA $\epsilon_1$ has positive chirality and $\epsilon_2$ has negative chirality. In type IIB both $\epsilon_{1,2}$ have positive chirality.

In this subsection, we will consider the geometry defined by \emph{one} Majorana--Weyl spinor $\epsilon$. We will work in a basis where all the $\gamma^M$ are real; the Majorana condition then simply means that $\epsilon$ is real. In frame indices, $\gamma^0$ is antisymmetric, whereas $\gamma^1,\ldots,\gamma^9$ are symmetric. This can be summarized by saying 
\begin{equation}\label{eq:gammat}
	\gamma^t_M = \gamma^0 \gamma_M \gamma^0 \ ;
\end{equation}
in other words, $\gamma^0$ is the intertwiner between the representations $\{\gamma_M\}$ and $\{\gamma^t_M\}$ of the Clifford algebra.

\subsubsection{Forms defined by $\epsilon$} 
\label{ssub:forms}

We will start with some preliminaries on how spinors are related to differential forms in ten dimensions. As usual, to a differential form we can associate a bispinor via the Clifford map: 
\begin{equation}\label{eq:cliffordmap}
	C_k\equiv \frac1{k!} C_{M_1\ldots M_k} dx^{M_1}\wedge \ldots \wedge dx^{M_k} \ \longrightarrow \ \sla C_k \equiv \frac1{k!} C_{M_1\ldots M_k} \gamma^{M_1\ldots M_k}\ .
\end{equation}
Many formulas about bispinors are usefully summarized by this notation. For example, we will need in what follows:
\begin{align}
	\label{eq:star}
	&\gamma \,\sla C_k = \slash{6}{40}{* \lambda(C_k)} \ ,\\
	\label{eq:gCg}
	& \gamma^M \sla C_k \gamma_M = (-)^k (10-2k) \sla C_k 
\end{align}
where $\gamma= \gamma^0 \gamma^1 \ldots \gamma^9$ is the chiral operator, and $\lambda(C_k)\equiv (-1)^{\lfloor \frac k2 \rfloor} C_k$. The generalizations of (\ref{eq:star}), (\ref{eq:gCg}) to any dimension (which we will need in section \ref{sec:examples}) can be found in (\ref{eq:stard}) and (\ref{eq:gCgd}) in appendix \ref{sec:bispinors}. From now on, we will drop the slash, and freely confuse differential forms with the associated bispinors. It is also useful to recall how wedges and contractions are related to Clifford products: 
\begin{equation}\label{eq:gwedge}
	\gamma^M  C_k = (dx^M\wedge + \iota^M)  C_k
	\ ,\qquad
	C_k \gamma^M = (-)^k (dx^M \wedge - \iota^M) C_k
\end{equation}
where $\iota^M \equiv g^{MN}\iota_N \equiv g^{MN}\iota_{\del/\del x^N}$. We will also sometimes use the notation
\begin{equation}\label{eq:garrow}
	\stackrel\to \gamma_M = (dx_M + \iota_M)\ ,\qquad
	\stackrel \leftarrow \gamma_M = (dx_M - \iota_M)(-)^{\rm deg} \ ,
\end{equation}
where $(-)^{\rm deg} C_k \equiv (-)^k C_k$, on a $k$-form $C_k$.

Consider now a Majorana--Weyl $\epsilon$ of chirality $\pm 1$. 	We can build from it the bispinor $\epsilon \otimes \overline{\epsilon}$, where 
\begin{equation}\label{eq:epsbar}
	\overline{\epsilon}\equiv \epsilon^t \gamma^0\ .
\end{equation}
The Fierz identities (see (\ref{eq:fierzd})) allow us to rewrite it as a sum of differential forms: 
\begin{equation}\label{eq:ee}
	\epsilon \otimes \overline{\epsilon}= 
	\sum_k \frac1{32 k!}(\overline{\epsilon}\,\gamma_{M_k \ldots M_1}\epsilon) \gamma^{M_1 \ldots M_k}\ .
\end{equation}	
Actually, most of the bilinears $\overline{\epsilon} \gamma_{M_1 \ldots M_k}\epsilon$ vanish identically. First of all, $\gamma^0 \gamma_{M_1 \ldots M_k}$ is an antisymmetric matrix for $k=0,3,4,7,8$, so the corresponding bilinears vanish. Secondly, since $\overline{\epsilon}$ has chirality opposite to $\epsilon$, the bilinears vanish when $k$ is even. That leaves us with three cases: $k=1,5,9$. The case $k=9$ is $*$-dual to the case $k=1$, thanks to (\ref{eq:star})\footnote{Notice also that $*\Omega_5= \pm \Omega_5$.}; so the independent bilinears are 
\begin{equation}\label{eq:KM}
K_M \equiv \frac1{32} \overline{\epsilon} \gamma_M \epsilon
\ ,\qquad
\Omega_{M_1\ldots M_5} \equiv \frac1{32} \overline{\epsilon} \gamma_{M_1\ldots M_5} \epsilon\ ,
\end{equation}
and (\ref{eq:ee}) reads
\begin{equation}\label{eq:KO}
\epsilon \otimes \overline{\epsilon} = 
K + \Omega_5 \pm * K = K (1 \mp \gamma) + \Omega_5 \ . 
\end{equation}

We can now compute, using (\ref{eq:gCg}), 
\begin{equation}\label{eq:geeg}
	\gamma^M \epsilon \otimes \overline{\epsilon}\, \gamma_M = 
	-8 K (1 \pm \gamma) \ ;
\end{equation}
it follows that
\begin{equation}\label{eq:Ke}
	K \epsilon = K_M \gamma^M \epsilon= \frac1{32}\gamma^M \epsilon\,\overline{\epsilon}\gamma_M \epsilon = -\frac14 K (1\pm \gamma)\epsilon = -\frac12 K \epsilon \ \ \Rightarrow \ \  
	K \epsilon = 0 \ . 
\end{equation}
(Recall that $\epsilon$ has chirality $\pm1$, as declared before (\ref{eq:epsbar}).) Hitting this result from the left by $\overline{\epsilon}$, we get
\begin{equation}\label{eq:Knull}
	K^M K_M = 0 \ .
\end{equation}
So $K^M$ is a null vector.


\subsubsection{Structure group from gamma matrices} 
\label{ssub:structure}

We will now determine the structure group defined by $\epsilon$, which is the stabilizer (or isotropy group, or little group) for the ${\rm Spin}(9,1)$ action on it. We will show that the structure group is ${\rm Spin}(7)\ltimes \rr^8$, and that the orbit of the action is 16-dimensional. (A similar computation can be found in \cite{bryant-spinors}, and more explicitly in \cite{figueroaofarrill}.)

Given a ten-dimensional Majorana--Weyl spinor $\epsilon$, we have seen in section \ref{sec:gen} that the bilinear $K_M$ in (\ref{eq:KM}) is a null vector. We will assume in what follows that $\epsilon$ has chirality $+$; the discussion is virtually the same for chirality $-$.

We will choose a frame in which this vector is the vielbein $e_-$: 
\begin{equation}
	K= e_-\ .
\end{equation}
Since $K$ is null, there are eight more vectors which are orthogonal to it; choose a basis $e_\alpha$ for them, $\alpha=1,\ldots,8$. Finally, we have to pick one more direction, $e_+$, which is \emph{not} orthogonal to $K$; we will take
\begin{equation}
	e_\pm \cdot e_\pm = 0 \ ,\qquad e_- \cdot e_+ = \frac12 \ ,\qquad 
	e_\pm \cdot e_\alpha = 0 \ . 
\end{equation}

The gamma matrices in this frame are $\gamma_-=K\cdot$, $\gamma_+$, $\gamma_\alpha$. So for example (\ref{eq:Ke}) reads
\begin{equation}\label{eq:g+e}
	\gamma^+ \epsilon= 0= \gamma_- \epsilon \ .
\end{equation}
Our decomposition of indices suggests to pick a basis for these gamma matrices where 
\begin{equation}
	\gamma_\pm = \gamma^{\pm}_{(2)}\otimes 1_8 \ ,\qquad \gamma^\alpha= (\gamma^{+-}_{(2)})\otimes \gamma^\alpha_{(8)} \ ,
\end{equation}
where $\{\gamma^\pm_{(2)}\}$ and $\{\gamma^\alpha_{(8)}\}$ are bases for the two- and eight-dimensional Clifford algebras respectively. In fact, we will take $\gamma_{(2)\,-}={{0\ 1}\choose {0 \ 0}}$, $\gamma_{(2)\,+}={{0\ 0}\choose {1 \ 0}}$, so that, from $K\cdot\epsilon= \gamma_- \epsilon=0$, it follows that
\begin{equation}
	\epsilon= | \uparrow \ \rangle \otimes \eta_+\ ,\qquad | \uparrow\ \rangle \equiv \left(\begin{array}{c}
		1 \\ 0 
	\end{array}\right)\ ,
\end{equation}
where $\eta_+$ is an eight-dimensional Majorana--Weyl spinor. 

We know from (\ref{eq:g+e}) that one gamma matrix annihilates $\epsilon$; a priori there could be more. In the basis we have chosen, this could come from an eight-dimensional gamma matrix: $\gamma^\alpha_{(8)} \eta_+ \stackrel ?= 0$. 
But a Majorana--Weyl spinor in eight dimensions is not annihilated by any linear combination of gamma matricess. So $\epsilon$ is not annihilated by any vector other than $K= \gamma_-$. This makes it very different from a pure spinor, which would be annihilated by five gamma matrices. A consequence of this will be that our ``generalized'' treatment of the ten-dimensional supersymmetry conditions will not deal with pure spinors, unlike the treatment of flux compactifications in \cite{gmpt2,gmpt3}.

We can now look at the infinitesimal action of a Lorentz transformation on $\epsilon$:
\begin{equation}
	\delta \epsilon = \omega_{AB}\gamma^{AB} \epsilon\ .
\end{equation}
We have to ask which products of two gamma matrices annihilate $\epsilon$. One obvious such product is $\gamma^{+}\gamma^\alpha= \gamma^{+\alpha}$, since already $\gamma^+$ annihilates $\epsilon$.
This time, however, we also find a contribution from the eight dimensional gamma matrices, since a Majorana--Weyl $\eta_+$ is annihilated by 21 out of 28 of the $\gamma^{\alpha \beta}_{(8)}$. Group theoretically, the representation \textbf{28} of SO(8) decomposes as $\mathbf{21}\oplus \mathbf{7}$ under its subgroup Spin(7). So we can write:
\begin{equation}\label{eq:stabeps}
	{\rm stab}(\epsilon)= {\rm span}\{ \omega_{\bf 21}^{\alpha \beta}\, \gamma_{\alpha \beta}\ , 
	\gamma^{+ \alpha} \}\ ,
\end{equation}
where $\omega_{\bf 21}^{\alpha \beta}$ is any two-form in the \textbf{21} of Spin(7). Notice that this is the adjoint; so the $\omega_{\bf 21}^{\alpha \beta}\, \gamma_{\alpha \beta}$ generate the Lie algebra of Spin(7). Moreover, their commutation relations with the $\gamma^{+	\alpha}$ are those of the semi-direct group
\begin{equation}\label{eq:ISpin7}
	{\rm ISpin}(7)\equiv {\rm Spin}(7)\ltimes \rr^8\ ;
\end{equation}
we have introduced the notation ISpin, for ``inhomogeneous Spin'', in analogy to the notation ${\rm ISO}(d)$ for inhomogeneous ${\rm SO}(d)$ groups.

We can now also look at the orbit of the Lorentz group action, which is given by all spinors that can be written as $\gamma^{AB}\epsilon$. We can already determine the dimension of this space as the dimension of  ${\rm Spin}(9,1)$ minus the dimension of the isotropy group (\ref{eq:ISpin7}): this gives 45-29=16. So we expect the orbit to be 16-dimensional. Let us see this more explicitly.  
The only subtle components are the purely eight-dimensional ones, $\gamma^{\alpha \beta} \eta_+$. We have just seen that the contribution from the \textbf{21} vanishes; so only the contribution from the \textbf{7} is non-zero:
\begin{equation}\label{eq:epsorbit}
	\omega_\mathbf{7}^{\alpha \beta} \,\gamma_{\alpha \beta}\epsilon= | \uparrow \ \rangle \otimes (\Pi_\mathbf{7}^{\alpha \beta}{}_{\gamma \delta})\,\gamma_{(8)}^{\gamma \delta}\eta_+\ , \qquad
	\gamma^{-\alpha} \epsilon = 2 \,| \downarrow \ \rangle \otimes \gamma_{(8)}^\alpha \eta_+\ ,\qquad
	\gamma^{+-} \epsilon =2 \epsilon \ .
\end{equation}
(We have used our normalization $g^{+-}=2$.) So we have $7+8+1=16$ non-zero elements that can be obtained from the Lorentz infinitesimal action. This confirms that the orbit of $\epsilon$ is 16-dimensional.

It can in fact be shown \cite{bryant-spinors} that the action of the Lorentz group on the space $\Sigma_\pm$ of Majorana--Weyl spinors of either chirality has only two orbits: the zero spinor, and everything else. In other words, any two non-vanishing spinors of the same chirality can be mapped to one another by a Lorentz transformation. All non-vanishing Weyl spinors have then the same  stabilizer.


\subsubsection{Structure group from forms} 
\label{ssub:structureforms}

	The structure group ${\rm Spin}(7)\ltimes \rr^8$ can also be understood from the point of view of the forms $K$, $\Omega_5$ defined in (\ref{eq:KM}). Namely, one can find it as the stabilizer of these two forms for the action of the Lorentz group ${\rm SO}(9,1)$.
	
	We start by computing the stabilizer of $K$. Since $K$ is null: 
\begin{equation}\label{eq:ISO8}
		{\rm stab}(K)={\rm ISO}(8)={\rm SO}(8)\ltimes \rr^8\ .
\end{equation}	
This is just the generalization of the familiar little group ISO(2) of a null vector in four dimensions (see for example \cite[Ch.~2.5]{weinberg}); in that case, the quantum number of the ${\rm SO}(2)$ part of ${\rm ISO}(2)$ is helicity. 

	We now have to ask which subgroup of ISO(8) keeps also $\Omega_5$ invariant. Notice that (\ref{eq:Ke}) implies $K (\epsilon \otimes \overline{\epsilon})=0=(\epsilon \otimes \overline{\epsilon}) K$. Using (\ref{eq:gwedge}), this implies $K\wedge (\epsilon \otimes \overline{\epsilon})= \iota_K (\epsilon \otimes \overline{\epsilon})=0$. Recalling (\ref{eq:KO}), we get: 
\begin{equation}
	K\wedge \Omega_5 = \iota_K \Omega_5 = 0 \ .
\end{equation} 
This implies 
\begin{equation}\label{eq:O4}
	\Omega_5 = K \wedge \Psi_4
\end{equation}
for some four-form $\Psi_4$. The form $\Psi_4$ can also be understood as follows: consider the nine-dimensional space $K^\perp$ of vectors orthogonal to $K$. Since $K$ is null, $K\in K^\perp$. Then we can define the quotient
\begin{equation}\label{eq:K8}
	K_8 \equiv K^\perp/\langle K \rangle
\end{equation}
 of vectors which are orthogonal to $K$, modulo vectors which are proportional to $K$. If we restrict our original spinor $\epsilon$ to $K_8$, we obtain a Majorana--Weyl spinor in eight dimensions; this is known to give rise to a ${\rm Spin}(7)$ structure. In fact $\Psi_4$ in (\ref{eq:O4}) is nothing but the four-form that describes this Spin(7) structure. If in the little group of $K$,  ${\rm stab}(K)={\rm ISO}(8)$, we consider the transformations that also leave this Spin(7) structure invariant, we reduce the SO(8) factor to Spin(7). This gives an alternative understanding to the stabilizer~(\ref{eq:ISpin7}). 

We can also now notice that the map 
\begin{equation}\label{eq:hopf}
		\epsilon \mapsto K
\end{equation}
is a Hopf fibration. The space of $\epsilon$ such that $\epsilon^t \epsilon=1$ is a sphere $S^{15}$.  (\ref{eq:hopf}) maps this to the space of $K$ which are null and such that $K_0=1/32$. This is a slice of the light cone, so it is a copy of $S^8$. The fibre of the map (\ref{eq:hopf}) is then the space of $\epsilon$'s that map to the same $K$: this is ${\rm Spin}(8)/{\rm Spin}(7)\cong S^7$. All this can be made more transparent by using an octonion basis for the gamma matrices, as for example in \cite{bryant-orbits}; $S^8$ is then understood as the octonionic projective line ${\Bbb O}\pp^1$.



\subsection{Geometry defined by two spinors} 
\label{sub:epseps}

	We will now move on to considering what happens with two different spinors $\epsilon^1$ and $\epsilon^2$, which is what we need in type II theories. As we will see, there are various possibilities for the structure group in $T$, whereas the structure group defined in $T \oplus T^*$ is universal. This is similar to what one finds for four-dimensional vacua \cite{gmpt2}. In that case, one finds both SU(2) and SU(3) structures on $T$, and ${\rm SU}(3) \times {\rm SU}(3)$ on $T \oplus T^*$. 
		
	The list of structure groups in section \ref{ssub:epsepsstr} in itself is a curiosity; it will be a useful preliminary, however, towards writing down the possible explicit expressions for $\Phi$ in section \ref{ssub:epsepsforms}. Moreover, the list of the generators in the stabilizer will be useful when we ask whether $\Phi$ determines a metric in section \ref{sub:metric}. 
		
\subsubsection{Structure groups} 
\label{ssub:epsepsstr}

We have seen in section \ref{sub:eps} that a single ten-dimensional spinor defines an ISpin(7)$\equiv {\rm Spin}(7)\ltimes \rr^8$ structure. With two spinors, we have to consider the isotropy group in SO(9,1) of both $\epsilon_{1,2}$. This is the intersection of two copies of ISpin(7); there are various possibilities, which have been listed for example in \cite{gran-gutowski-papadopoulos,gran-gutowski-papadopoulos-2}. We will now give a quick description of the various cases; we will describe them in more detail in section \ref{ssub:epsepsforms}.

In IIA, $\epsilon_1$ and $\epsilon_2$ have opposite chiralities. If the two null vectors $K_1$ and $K_2$ defined by them are proportional, we can use the gamma matrix basis defined in section \ref{ssub:structure} for both of them. We then reduce ourselves to considering two eight-dimensional spinors of opposite chirality. The intersection of their stabilizers is $G_2$. So overall we have a $G_2 \ltimes \rr^8$ structure. If, on the other hand, the two null vectors $K_1$ and $K_2$ are not proportional, without loss of generality we can assume that they are respectively $e^+$ and $e^-$ (up to a rescaling). The two spinors can then be written as $\epsilon_1= | \!\uparrow\, \rangle \otimes \eta_1 = {\eta_1 \choose 0}$ and $\epsilon_2= | \!\downarrow\, \rangle \otimes \eta_2 = {0 \choose \eta_2}$. We are then reduced to the common stabilizer of the two eight-dimensional spinors $\eta_1$ and $\eta_2$, which have the same chiralities. This is generically SU(4), but can get enhanced to Spin(7) if $\eta_1$ and $\eta_2$ are proportional. So we have found three possibilities: 
\begin{equation}\label{eq:epsepsIIA}
	G_2 \ltimes \rr^8 \ ,\qquad {\rm SU}(4) \ ,\qquad {\rm Spin}(7)\qquad \qquad
	({\rm on}\ T;\ {\rm in\ IIA})\ .
\end{equation}
Before we move on to IIB, it is interesting to compare (\ref{eq:epsepsIIA}) with what happens \cite{bryant-orbits,gauntlett-pakis,gauntlett-gutowski-pakis} in \emph{eleven} dimensions. There is a single supersymmetry parameter $\epsilon$, which defines a vector $K^M_{11} = \overline{\epsilon_{11}} \gamma^M \epsilon_{11}$. However, unlike our ten-dimensional $K_{1,2}$, which are both always null, $K_{11}$ can be either timelike or null; even if the component of $K_{11}$ along $x^{10}$ vanishes, there is no contradiction, since its projection along the remaining ten dimensions is $K_1+K_2$, and the sum of two null vectors can be either timelike or null. The little group of $\epsilon_{11}$ is SU(5) when $K_{11}$ is timelike and $({\rm Spin(7)\ltimes \rr^8})\times \rr$ when $\epsilon_{11}$ is null. When $K_{11}$ is timelike, $K_1$ and $K_2$ cannot be proportional, and we get SU(4) in ten dimensions. When $K_{11}$ is null, $K_1$ and $K_2$ can be either proportional (in which case we get $G_2 \ltimes \rr^8$ in ten dimensions) or not (in which case we get ${\rm Spin}(7)$).

Coming now to IIB, $\epsilon_1$ and $\epsilon_2$ have the same chirality. If the two null vectors $K_1$ and $K_2$ are proportional, again we can use the gamma matrix basis defined in section \ref{ssub:structure} for both, and we can write $\epsilon_i = | \! \uparrow\,  \rangle \otimes \eta_i$, where $\eta_i$ are eight-dimensional spinors of the same chirality. The intersection of the stabilizers of the $\eta_i$ is generically SU(4), but can get enhanced to Spin(7) if they are proportional. So we conclude that the common stabilizer of the $\epsilon_i$ is generically ${\rm SU}(4)\ltimes \rr^8$, and ${\rm Spin}(7)\ltimes \rr^8$ when $\epsilon_1$ and $\epsilon_2$ are proportional. When $K_1$ and $K_2$ are not proportional, again without loss of generality we can assume that they are respectively $e^+$ and $e^-$ (up to a rescaling). The two spinors can then be written as $\epsilon_1= | \!\uparrow\, \rangle \otimes \eta_1 = {\eta_1 \choose 0}$ and $\epsilon_2= | \!\downarrow\, \rangle \otimes \eta_2 = {0 \choose \eta_2}$, where this time $\eta_1$ and $\eta_2$ have opposite chiralities. The common stabilizer of two eight-dimensional spinors with opposite chiralities is $G_2$. In conclusion, we have found three possibilities: 
\begin{equation}\label{eq:epsepsIIB}
	{\rm SU}(4)\ltimes \rr^8 \ ,\qquad {\rm Spin}(7)\ltimes \rr^8 
	\ ,\qquad G_2\qquad\qquad ({\rm on}\ T;\ {\rm in\ IIB})\ .
\end{equation}

The occurrence of all these cases is similar to the appearance of both SU(2) and SU(3) structures in the classification of type II \emph{vacua}, namely solutions of the form $\rr^{1,3}\times M_6$ or AdS$_4\times M_6$. Using the differential geometry associated with the  structure groups in~(\ref{eq:epsepsIIA}) and~(\ref{eq:epsepsIIB}) would be complicated, and it would give rise to a plethora of ``intrinsic torsion'' classes. Moreover, the stabilizer of the spinors $\epsilon_i$ may change from a point to another, even for a single solution.

In the case of vacua, the classification is more elegant \cite{gmpt2} when one considers the structure group in $T\oplus T^*$: one obtains there an ${\rm SU}(3) \times {\rm SU}(3)$ structure. In the same spirit, we will now show that in all these cases one can define the same structure group on $T \oplus T^*$, using the single bispinor
\begin{equation}\label{eq:Phi}
	\Phi= \epsilon_1 \otimes \overline{\epsilon_2}\ .
\end{equation}
One might also think of considering $\epsilon_1 \otimes \overline{\epsilon_1}$ or $\epsilon_2 \otimes \overline{\epsilon_2}$. As we will show in section \ref{sec:diff}, however, considering $\Phi$ in (\ref{eq:Phi}) (along with some descendants that we will introduce shortly) is enough to recast the conditions for unbroken supersymmetry in geometrical language. Notice that $\Phi$ is not a pure spinor. To see this, we can use the gamma matrix basis described in section \ref{ssub:structure}. Notice, however, that that basis will in general be different for the spinors $\epsilon_1$ and $\epsilon_2$. To take care of that, we will add subscripts ${}_1$ or ${}_2$ to all indices (similarly to the notation used in \cite[App.~A.4]{gmpt3} to distinguish the two almost complex structures defined by a pure spinor pair). In this notation, the only sections of $T \oplus T^*$ that annihilate $\Phi$ are
\begin{equation}\label{eq:-1-2}
	{\rm Ann}(\Phi)= {\rm span}\{\stackrel \to \gamma_{-_1} \ , \stackrel \leftarrow \gamma_{-_2}\} \ ,
\end{equation}	
whereas a pure spinor would have an annihilator of dimension 10. So $\Phi$ is quite different in character from the forms $\phi_\pm$ that can be used to reformulate the supersymmetry conditions for four-dimensional vacua \cite{gmpt2,gmpt3}.

The bundle $T \oplus T^*$ has rank 20, and it has a natural metric ${\cal I}\equiv {{0 \ 1}\choose {1 \ 0}}$ defined by contracting one-forms with vectors. Its structure group is then ${\rm SO}(10,10)$. To see how this structure group is reduced by $\Phi$, we need to compute the stabilizer of $\Phi$ inside ${\rm SO}(10,10)$. The infinitesimal action of SO(10,10) is given by operators of the form 
\begin{equation}\label{eq:so1010}
	\omega_{AB} \Gamma^{AB}\ ,
\end{equation}
where $\Gamma^A=\{ dx^m \wedge , \iota_m \}$, which generate the Clifford algebra Cl(10,10). The computation is much easier if one changes basis, using (\ref{eq:garrow}), to the ordinary Cl(9,1) gamma matrices acting from the left and from the right on a bispinor. We get:
\begin{equation}\label{eq:stabPhi}
	{\rm stab}(\Phi)= {\rm span}\left\{
	\begin{array}{c}
			\omega_{\bf 21}^{\alpha_1 \beta_1} \stackrel\to\gamma_{\alpha_1 \beta_1}\, ,\ \stackrel \to \gamma_{-_1 \alpha_1}\, , \ \omega_{\bf 21}^{\alpha_2 \beta_2} \stackrel\leftarrow\gamma_{\alpha_2 \beta_2}\, , \ \stackrel\leftarrow \gamma_{-_2 \alpha_2}\, , \ 
			\stackrel \to \gamma_{+_1 -_1} + \stackrel \leftarrow \gamma_{+_2 -_2}
			\vspace{.1cm}\\ 
		\stackrel \to \gamma_{-_1} \stackrel \leftarrow \gamma_{\alpha_2}\, , \ 
		\stackrel \to \gamma_{-_1} \stackrel \leftarrow \gamma_{+_2}\, , \ 	
		\stackrel \to \gamma_{\alpha_1} \stackrel \leftarrow \gamma_{-_2}\, , \ 
		\stackrel \to \gamma_{+_1} \stackrel \leftarrow \gamma_{-_2}\, , \ 
		\stackrel \to \gamma_{-_1} \stackrel \leftarrow \gamma_{-_2}
	\end{array}
 \right\}\ .
\end{equation}
We have again used the notation, introduced above (\ref{eq:-1-2}), of adding an extra subscript $_1$ and $_2$ to indices relative to spinors $\epsilon_1$ and $\epsilon_2$ respectively.
The first line in (\ref{eq:stabPhi}) contains the stabilizers of $\epsilon_1$ and of $\overline{\epsilon_2}$. The last element of the first line comes about because $\gamma_{+-}\epsilon= -2 \epsilon$ (which just follows from Clifford algebra). The generators on the second line do not correspond to acting on the spinors; as we will see in section \ref{sub:metric}, they correspond to acting on the metric and $B$ field. 

Using the ordinary Cl(9,1) gamma matrix algebra, we can also determine the Lie algebra of ${\rm stab}(\Phi)$. To perform this computation, it is actually best to decorate again $\stackrel\leftarrow\gamma_M$ with a degree operator $(-)^{\rm deg}$, as in (\ref{eq:garrow}), so that
\begin{equation}
	\{ \stackrel \to \gamma_M , \stackrel \leftarrow \gamma_N(-)^{\rm deg}\} = 0 \ .
\end{equation}
The $\stackrel \to \gamma_M$ and $\stackrel \leftarrow \gamma_N (-)^{\rm deg}$ then generate two anticommuting copies of Cl(9,1). We now see that the generators $\omega_{\bf 21}^{\alpha_1 \beta_1} \stackrel\to\gamma_{\alpha_1 \beta_1}$ and $\omega_{\bf 21}^{\alpha_2 \beta_2} \stackrel\leftarrow\gamma_{\alpha_2 \beta_2}$ generate two copies of Spin(7). Another subalgebra is spanned by the three generators $\{\stackrel \to \gamma_{+_1 -_1} + \stackrel \leftarrow \gamma_{+_2 -_2}\,,\ \stackrel \to \gamma_{-_1} \stackrel \leftarrow \gamma_{+_2}(-)^{\rm deg}\,,\ \stackrel \to \gamma_{+_1} \stackrel \leftarrow \gamma_{-_2} (-)^{\rm deg} \}$; this is isomorphic to ${\rm Sl}(2,\rr)$. The remaining 33 generators satisfy the commutation relations of a Heisenberg algebra $H_{33}$, with $\stackrel \to \gamma_{-_1\alpha_1}$, $\stackrel \to \gamma_{-_1}\stackrel \leftarrow \gamma_{\alpha_2} (-)^{\rm deg}$ playing the role of the $x^I$, the $\stackrel \to \gamma_{\alpha_1} \stackrel \leftarrow \gamma_{-_2}(-)^{\rm deg}$, $\stackrel \leftarrow \gamma_{-_2 \alpha_2}$, playing the role of the $p_I$, and $\stackrel \to \gamma_{-_1} \stackrel \leftarrow \gamma_{-_2} (-)^{\rm deg}$ as the central element. Having divided the generators of ${\rm stab}(\Phi)$ in three subalgebras, we have to look at how these commute with each other; we find once again a semidirect product:
\begin{equation}\label{eq:genISpin}
	({\rm Spin}(7)^2 \times {\rm Sl}(2,\rr))\ltimes H_{33}\qquad\qquad({\rm on}\ T \oplus T^*;\ {\rm in\ IIA/IIB})\ .
\end{equation}
This is the structure group defined by $\Phi$ on $T \oplus T^*$. Since the group (\ref{eq:genISpin}) is a bit of a tongue-twister, we will simply say that $\Phi$ defines a \emph{generalized ${\rm ISpin}(7)$ structure}.


\subsubsection{Forms} 
\label{ssub:epsepsforms}

The fact that $\Phi$ defines a certain structure group on $T \oplus T^*$ will be important in reformulating the supersymmetry equations in terms of forms. However, in practice one also needs to know the possible explicit expressions for $\Phi$. There are several cases, corresponding to the various structure groups in $T$ we found in (\ref{eq:epsepsIIA}) and (\ref{eq:epsepsIIB}). 

As a preliminary, we will deal with the bilinears one gets in eight Euclidean dimensions. For simplicity, we can use a real basis for the gamma matrices. (A particularly nice one, which we mentioned earlier, can be written in terms of octonions; see for example \cite{bryant-orbits}.) Our bilinears will then all be real. 

We start with two real spinors $\eta_1$, $\eta_2$ of the same chirality. When $\eta_1$ and $\eta_2$ are proportional, the structure group is just the stabilizer of a real Weyl spinor in eight dimensions, which is Spin(7). In this case, the bilinear simply reads
\begin{equation}\label{eq:etaetaprop}
	\phi_{{\rm Spin}(7)}\equiv\eta_1 \otimes \eta_2^t = A(1+ \Psi_4 + {\rm vol}_8)\ , 
	\qquad \qquad (\eta_2 = A \eta_1)
\end{equation}
where $\Psi_4$ is the four-form that defines the Spin(7) structure. Notice that the two-form and six-form parts are absent: this follows from the fact that our gamma matrices are symmetric. 

When $\eta_1$ and $\eta_2$ are not proportional, they define generically an SU(4) structure, as we mentioned in section \ref{ssub:epsepsstr}. Let us review why. In general, an ${\rm SU}(d)$ structure is defined in $2d$ dimensions by a pure Weyl spinor. A Majorana spinor can never be pure, but we can combine our two spinors in $\eta \equiv \eta_1 + i \eta_2$, which is not Majorana, but still Weyl. However, in eight dimensions, not all Weyl spinors are pure: the space of pure spinors is $\cc \times {\rm SO}(8)/{\rm U}(4)$, which has real dimension 14. This is two less than 16, the real dimension of the space of all Weyl spinors. However, the constraint for a Weyl spinor $\eta$ to be pure is simply that\footnote{In general, purity is equivalent to the condition that all bilinears $\eta^t \gamma_{m_1 \ldots m_k} \eta$ should be zero except when $k$ is half the dimension of the space. In $d=8$, the cases $k=1,2,3$ vanish automatically, and we are left with the case $k=0$, which is (\ref{eq:etapure}).}
\begin{equation}\label{eq:etapure}
	\eta^t \eta=0\ .
\end{equation}
Now, any $\eta_1$ and $\eta_2$ which are not proportional can be parameterized as 
\begin{equation}\label{eq:etapar}
	\eta_1 = \cos(\psi) \tilde\eta_1 + \sin(\psi) \tilde\eta_2 \ ,\qquad
	\eta_2 = A (\cos(\psi)\tilde\eta_1 - \sin(\psi) \tilde\eta_2)\ ,
\end{equation}
	where $A$ and $\psi$ are real, and
\begin{equation}
	\tilde\eta_1^t \tilde\eta_2 = 0 \ ,\qquad \tilde\eta_1^t \tilde\eta_1 =  \tilde\eta_2^t \tilde\eta_2 \ .
\end{equation} 
We can now see that $\eta \equiv \tilde\eta_1 + i \tilde\eta_2$ satisfies (\ref{eq:etapure}), and hence it is pure. This shows that the two original spinors $\eta_1$ and $\eta_2$ define an SU(4) structure. 

We can now use this information to write down the bilinear $\eta_1 \otimes \eta_2^t$. Since $\eta$ is pure, its bilinears are simply\footnote{We have normalized $|| \eta ||^2=32$; this is no loss of generality for us, because we would be able in any case to reabsorb $|| \eta ||^2$ in the costant $A$ in (\ref{eq:etaeta}).}
\begin{equation}\label{eq:etaetapure}
	\eta \otimes \eta^t = \frac12 \Omega_4 \ ,\qquad
	\eta \otimes \eta^\dagger = \frac12 e^{iJ}\ , 
\end{equation}
where $\Omega_4$ and $J$ are simply the holomorphic and symplectic form associated to the SU(4) structure.
From (\ref{eq:etaetapure}) we can extract the bilinears $\tilde\eta_i \otimes \tilde\eta_j^t$; going back to (\ref{eq:etapar}) we obtain
\begin{equation}\label{eq:etaeta}
	\phi_{{\rm SU}(4)}\equiv\eta_1 \otimes \eta_2^t = A \, 
	{\rm Re} \Big( \Omega + e^{-2i \psi} e^{iJ}\Big)\ .
\end{equation}
The case where $\eta_1$ and $\eta_2$ are proportional is recovered as $\psi\to 0$. Indeed in this limit we see that the two-form and six-form parts disappear, and we are left with (\ref{eq:etaetaprop}), with $\Psi_4 = {\rm Re} \Omega- J^2/2$ being the four-form that defines the Spin(7) structure.

The other case to consider is when $\eta_1$ and $\eta_2$ have opposite chirality. In this case, one can define a vector $v_m = \frac1{16} \eta_2^t \gamma_m \eta_1$. Applying (\ref{eq:etaetaprop}) to the case $\eta_2=\eta_1$, one gets $\eta_2 \eta_2^t = 1 + \Psi_4 - {\rm vol}_8$ (the minus sign in front of ${\rm vol}_8$ being due to the fact that $\eta_2$ is now of negative chirality). Using (\ref{eq:gCgd}) one can now compute, similarly to (\ref{eq:Ke}):
\begin{equation}
	v \cdot \eta_2 = \frac1{16}\gamma^m \eta_2 \eta_2^t \gamma_m \eta_1 = \frac12 (1+ \gamma) \eta_1 = \eta_1\ ,
\end{equation}
from which we obtain
\begin{equation}\label{eq:etaetaG2}
	\phi_{G_2}\equiv\eta_1 \otimes \eta_2^t = v \cdot \eta_2 \eta_2^t = 
	v\cdot(1+ \Psi_4 - {\rm vol}_8)= v + \phi_3 + v\wedge \tilde \phi_4 - * v \ , \qquad \qquad (\eta_1 \, {\rm chir}. +; \ \eta_2 \, {\rm chir}. -)
\end{equation}
where $\phi_3\equiv v \llcorner \Psi_4$ (along with its seven-dimensional dual $\tilde\phi_4$) defines a $G_2$ structure. 

We can now use these eight-dimensional bilinears to compute the ten-dimensional ones. We can use the fact that the annihilator (\ref{eq:-1-2}) of $\Phi$ is generated by $\stackrel\to\gamma_{-_1}=K_1\wedge + K_1\llcorner$ and $\stackrel\leftarrow \gamma_{-_2}(-)^{\rm deg}= K_2\wedge - K_2\llcorner$. Actually, it will be convenient to work in terms of 
\begin{equation}\label{eq:KKt}
	K \equiv \frac12(K_1 + K_2) \ ,\qquad \tilde K \equiv \frac12(K_1 - K_2) \ ,
\end{equation}
so that for example
\begin{equation}\label{eq:KKtPhi}
	(\tilde K\wedge + K\llcorner) \Phi=0\ .
\end{equation}
By construction, $K\cdot \tilde K=0$, but in general neither $K$ nor $\tilde K$ is null. Their norms are $K^2=-\tilde K^2 = \frac12 K_1 \cdot K_2$, which is related in turn to $\Phi$ by 
\begin{equation}
	K_1 \cdot K_2 = \frac1{32} (-)^{{\rm deg}(\Phi)} (\Phi, \gamma^M \Phi \gamma_M)\ ,
\end{equation}
where we have used (\ref{eq:e2e1}), (\ref{eq:stard}) and (\ref{eq:mukaiTr}). $\gamma^M \Phi \gamma_M$ can be further evaluated using (\ref{eq:gCgd}).

When $K_1=K_2$, we see from (\ref{eq:KKtPhi}) that ${\rm Ann}(\Phi)$ contains $K\wedge$. So $\Phi$ should be of the form $K\wedge(\ldots)$. When $K_1\neq K_2$, $\Phi$ will be of the form $\exp\left[-\frac1{K^2}{K\wedge \tilde K}\right]\wedge (\ldots)$. The remaining parts $(\ldots)$ come from the eight-dimensional bilinears $\eta_1 \eta_2^t$ which we studied earlier. In IIA, the possible structure groups were listed in (\ref{eq:epsepsIIA}): we get
\begin{equation}\label{eq:PhiIIA}
	\begin{split}
		&\Phi_{G_2 \ltimes \rr^8}= K \wedge \phi_{G_2}\ ,\\
		&\Phi_{{\rm SU}(4)}= \exp\left[-\frac1{K^2}{K\wedge \tilde K}\right]\wedge \phi_{{\rm SU}(4)}\ , \qquad \qquad {\rm (IIA)}\\
		&\Phi_{{\rm Spin}(7)}= \exp\left[-\frac1{K^2}{K\wedge \tilde K}\right]\wedge \phi_{{\rm Spin}(7)}\ ,
	\end{split}
\end{equation}
where $\phi_{G_2}$ was given in (\ref{eq:etaetaG2}), $\phi_{{\rm SU}(4)}$ in (\ref{eq:etaeta}), and $\phi_{{\rm Spin}(7)}$ in (\ref{eq:etaetaprop}). In IIB, the possible structure groups were listed in (\ref{eq:epsepsIIB}), and we get
\begin{equation}\label{eq:PhiIIB}
	\begin{split}
		&\Phi_{{\rm SU}(4) \ltimes \rr^8}= K \wedge \phi_{{\rm SU}(4)}\ ,\\
		&\Phi_{{\rm Spin}(7) \ltimes \rr^8}= K\wedge \phi_{{\rm Spin}(7)}\ , \qquad \qquad {\rm (IIB)}\\
		&\Phi_{G_2}= \exp\left[-\frac1{K^2}{K\wedge \tilde K}\right]\wedge \phi_{G_2}\ .
	\end{split}
\end{equation}
In both IIA and IIB, as we recalled earlier, it is possible for $\Phi$ to be of a certain type at a point, and of a different type at another point.


\subsection{Spinor bilinears versus metric} 
\label{sub:metric}

In order to reformulate the conditions for supersymmetry in terms of $\Phi$, we need to know whether it determines a metric, partially or totally. In the case of the classification of vacua \cite{gmpt2}, the two pure spinors $\phi_\pm$ do determine a metric, and hence the supersymmetry equations can be rewritten as conditions on them and on nothing else. For our ten-dimensional generalization, we will see that $\Phi$ alone is not enough to determine a metric, so that some extra data are necessary. 

The way to determine whether a certain $G$-structure determines a globally defined metric $g_{MN}$ is to think of the latter as an ${\rm O}(d)$ structure. If $G$ is a subgroup of ${\rm O}(d)$, then the $G$-structure determines a metric. In abstract terms, this is because if the transition functions leave invariant the tensor $\omega$ defining the $G$-structure, they will also lie in ${\rm O}(d)$, and leave a metric invariant. In other words, if the stabilizer of $\omega$ is contained in ${\rm O}(d)$, it leaves invariant a quadratic form at every point; this quadratic form is $g_{MN}$.\footnote{\label{foot:id}Actually, if $G$ is contained in a smaller orthogonal group ${\rm O}(d')$, $d'<d$, then it is a subgroup of more than one copy of ${\rm O}(d)$, and the tensor $\omega$ will define more than one globally defined metric. For example, when $G$ is the trivial group (a so called ``identity structure''), the tangent bundle is trivial, and there is a basis of globally defined vectors.} On the other hand, if $G$ is \emph{not} a subgroup of ${\rm O}(d)$, then the stabilizer of $\omega$ contains an element that lies outside ${\rm O}(d)$, and that element can be used to change the metric without changing $\omega$; this shows that $\omega$ cannot determine a metric. 

We will now apply this general criterion to the case at hand; this will hopefully also make it clearer. Actually, it is more convenient to work on $T \oplus T^*$ and ask whether $\Phi$ determines a metric and $B$ field. This will spare us from having to go through all the cases in (\ref{eq:epsepsIIA}) and (\ref{eq:epsepsIIB}). We can use the fact that the data of $g_{MN}$ and $B_{MN}$ can be encoded in an ${\rm O}(9,1)\times {\rm O}(9,1)$ structure. This works as follows \cite[Chap.~6]{gualtieri}. An ${\rm O}(9,1)\times {\rm O}(9,1)$ structure is the definition of two rank 10 subbundles $C_\pm$ in $T \oplus T^*$; these can be singled out by a matrix ${\cal G}$ that is equal to $\mp 1_{10}$ on $C_\pm$ and hermitian with respect to the natural metric ${\cal I}={{0 \ 1}\choose {1\ 0}}$ on $T \oplus T^*$. From ${\cal G}$, one can write two orthogonal projections $\frac12(1\pm {\cal G})$ on $C_\pm$. It can be shown that such a ${\cal G}$ can be written as
\begin{equation}\label{eq:calG}
	{\cal G}={\cal E}^{-1}\left(\begin{array}{cc}
		-1&0\\0&1
	\end{array}\right){\cal E} =\left(\begin{array}{cc}
		-g^{-1} B & g^{-1}\\
		g-g^{-1} B g^{-1} & B g^{-1}
	\end{array}\right)\ ,\qquad 
	{\cal E}\equiv \left(\begin{array}{cc}
	1 & 1 \\
	E & -E^t	
	\end{array}\right)\ ,\qquad
	E\equiv g + B \ ,
\end{equation}
for some $g$ and $B$, which can be identified as the metric and $B$ field. So the data of $g$ and $B$ are encoded in a ${\rm O}(9,1)\times {\rm O}(9,1)$ structure, and we now need to ask whether the stabilizer of $\Phi$ defines a subgroup of ${\rm O}(9,1)\times {\rm O}(9,1)$. 

This is not intuitively obvious from the Lie algebra structure of ${\rm stab}(\Phi)$ from (\ref{eq:genISpin}). It is better to go back to the explicit expression (\ref{eq:stabPhi}). The advantage of this is that the gamma matrices acting from the left and from the right, (\ref{eq:garrow}), can be directly interpreted as belonging to the two bundles $C_\pm\subset T \oplus T^*$ that define the ${\rm O}(9,1)\times {\rm O}(9,1)$ structure. To see this, consider first the case $B=0$. We recognize that the matrix ${\cal E}$ in (\ref{eq:calG}) gives the change of basis in (\ref{eq:garrow}); in this basis, the fact that ${\cal G}= {\cal E}{{-1 \ 0}\choose {0\ \ 1 }} {\cal E}^{-1}$ just tells us that ${\cal G}$ gives $\mp 1$ on the gamma matrices acting from the left and from the right, respectively. So $C_+$ is the bundle of left-acting gamma matrices, and $C_-$ is the bundle of right-acting gamma matrices. When $B$ is non-zero, the change of basis ${\cal E}$ identifies $C_\pm$ as generated by 
\begin{equation}\label{eq:gammaE}
	e^{B\wedge} \stackrel \to \gamma_M e^{-B\wedge}= \iota_M + E_{MN}dx^N\wedge\ ,\qquad
	e^{B\wedge} \stackrel \leftarrow \gamma_M e^{-B\wedge} (-)^{\rm deg}= -\iota_M + E_{NM}dx^N\wedge\ .
\end{equation}
These again generate two copies of Cl(9,1). 

We see that for a bispinor of the form $\epsilon_1 \otimes \overline{\epsilon_2}$,  the infinitesimal generators of the Lie algebra of ${\rm O}(9,1)\times {\rm O}(9,1)$ should be 
\begin{equation}\label{eq:so912}
	{\rm so}(9,1)\oplus {\rm so}(9,1)= \{ \stackrel \to \gamma_{MN}\,,\ \stackrel \leftarrow \gamma_{MN}\}\ .
\end{equation}
(For a bispinor $e^B\wedge \epsilon_1 \otimes \overline{\epsilon_2}$, the gamma matrices in (\ref{eq:so912}) should be conjugated by $e^B$ as in~(\ref{eq:gammaE}).) However, we see that ${\rm stab}(\Phi)$ in (\ref{eq:stabPhi}) also contains elements of the type $\stackrel \to \gamma_M \stackrel \leftarrow \gamma_N$, which are not of the form (\ref{eq:so912}). This already tells us that ${\rm stab}(\Phi)$ is not a subgroup of ${\rm O}(9,1)\times {\rm O}(9,1)$. Hence $\Phi$ should not determine a metric and $B$ field. 

Let us try to get a more concrete understanding of why elements of the form $\stackrel \to \gamma_M \stackrel \leftarrow \gamma_N$ signal that $\Phi$ does not determine the metric and $B$ field. From (\ref{eq:garrow}) we can compute
\begin{subequations}
	\begin{align}
		\label{eq:dxi}
		&dx_M \wedge \iota_N = 
		\frac12\left( - \stackrel \to \gamma_{(M} \stackrel \leftarrow \gamma_{N)}(-)^{\rm deg}+ g_{MN}\right) + \frac14\left(\stackrel \to \gamma_{MN}+ \stackrel \leftarrow \gamma_{MN}\right)\ , \\
		\label{eq:dxdx}
		&dx^M \wedge dx^N\wedge = 
		 \frac12\stackrel \to \gamma_{[M} \stackrel \leftarrow \gamma_{N]}(-)^{\rm deg}+\frac14\left(\stackrel \to \gamma_{MN}- \stackrel \leftarrow \gamma_{MN}\right)\ .
	\end{align}
\end{subequations}
The symmetric part of (\ref{eq:dxi}) gives 
\begin{equation}
	dx_{(M} \wedge \iota_{N)} = 
	\frac12\left( - \stackrel \to \gamma_{(M} \stackrel \leftarrow \gamma_{N)}(-)^{\rm deg}+ g_{MN}\right)\ .
\end{equation}
We can interpret this as the effect on a bispinor of a change in metric. This comes about because the Clifford map (\ref{eq:cliffordmap}) depends on the metric, through the gamma matrices $\gamma^M$. If we deform the vielbeine as 
\begin{equation}
	\delta e^A_M = \frac12 \beta_M{}^N e^A_N\ ,
\end{equation}
the inverse vielbein transforms as $\delta e_A^M = -\frac12 \beta^M{}_N e_A^N$, the gamma matrices as $\delta \gamma^M = -\frac12 \beta^M{}_N \gamma^N$, and the metric as
\begin{equation}
	\delta g_{MN}= 2 e_{A(M}\delta e^A_{N)}= \beta_{(MN)}\ .
\end{equation}
We can then take $\beta_{MN}$ to be symmetric. The Clifford map is deformed as
\begin{equation}
	\delta \sla C_k= -\frac12 \beta_N{}^M 
	\slash{6}{60}{dx^N\wedge \iota_M C_k} \ .
\end{equation}
So the operator that takes into account the change in metric on a bispinor is 
\begin{equation}\label{eq:deltacliff}
	-\frac12 \delta g_{MN} dx^M\wedge \iota^N= -\frac14\left( - \stackrel \to \gamma^{(M} \stackrel \leftarrow \gamma^{N)}(-)^{\rm deg}+ g^{MN}\right)\delta g_{MN}\ .
\end{equation}


A change in the $B$ field, on the other hand, is simply given by $\Phi \to e^B\wedge \Phi$. Infinitesimally, (\ref{eq:dxdx}) shows us that this is given by $\stackrel \to \gamma_{[M} \stackrel \leftarrow \gamma_{N]}$ together with an action on the spinors $\epsilon_1$ and $\epsilon_2$ in $\Phi=\epsilon_1 \otimes \overline{\epsilon_2}$. 

So we see that, if we have elements of the form $\stackrel \to \gamma_M \stackrel \leftarrow \gamma_N$, they can be interpreted as a change in metric and $B$ field that does not change $\Phi$. Since our $\Phi$ does have such elements in its stabilizer (\ref{eq:stabPhi}), it cannot determine by itself a metric and $B$ field.

It is interesting to compare this failure with the way a pair of pure spinors $\phi_\pm$ determine the metric and $B$ field for vacua in \cite{gmpt2}. In that case, the common stabilizer of $\phi_+=\eta^1_+ \eta^{2\,\dagger}_+$ and $\phi_-=\eta^1_+ \eta^{2\,\dagger}_-$ is given by the union of the stabilizers of $\eta^1$ and $\eta^2$, ${\rm span}\{ \omega_{\bf 8}^{i_1 \bar j_1} \stackrel \to \gamma_{i_1 \bar j_1},$ $ \omega_{\bf 8}^{i_2 \bar j_2} \stackrel \leftarrow \gamma_{i_2 \bar j_2}\}$ (for more details, see \cite[App.~A.4]{gmpt3}). This stabilizer does not contain any elements of the type $\stackrel \to \gamma_M \stackrel \leftarrow \gamma_N$; in fact, it is isomorphic to ${\rm SU}(3) \times {\rm SU}(3)$, which is a subgroup of ${\rm SO}(6)\times {\rm SO}(6)$. In this case, $\phi_\pm$ do determine a metric and $B$ field.


\subsection{Adding vectors} 
\label{sub:vectors}


In section \ref{sub:metric}, we have found that a generalized ISpin(7) structure $\Phi$ does not determine uniquely a metric and $B$ field.  We will now try to add more degrees of freedom to $\Phi$, so as to resolve this ambiguity.

It is useful to start from a particular manifestation of the problem. Given an explicit form such as the ones we presented in section \ref{ssub:epsepsforms}, it is easy to compute a two-dimensional space ${\rm Ann}(\Phi)$ of sections of $T \oplus T^*$ that annihilate $\Phi$. We noticed in (\ref{eq:-1-2}) that this is given by the span of $\stackrel \to \gamma_{-_1}$ and $\stackrel \leftarrow \gamma_{-_2}$. With no additional information, however, we have no means of telling which element in this two-dimensional space is $\stackrel \to \gamma_{-_1}$ and which one is $\stackrel \leftarrow \gamma_{-_2}$.

So we should try to pick these two elements of ${\rm Ann}(\Phi)$. It is actually best to declare which elements of $T \oplus T^*$ correspond to the creators: 
\begin{equation}\label{eq:+1+2}
	\stackrel\to \gamma_{+_1}= e_{+_1}\cdot(\ )\ ,\qquad \stackrel \leftarrow \gamma_{+_2}(-)^{\rm deg}=(\ ) \cdot e_{+_2} \ .
\end{equation}
Here we introduced an alternative notation that will be useful later. $e_{+_1}$, $e_{+_2}$ are the vector parts of $\stackrel \to \gamma_{+_1}$ and $\stackrel \leftarrow \gamma_{+_2}$, and the symbol $\cdot$ denotes Clifford multiplication;  
(\ref{eq:+1+2}) are then two elements of $T \oplus T^*$. To formalize the fact that they are two creators, we should demand that
\begin{equation}\label{eq:comp}
	(e_{+_1}\cdot\Phi\cdot e_{+_2}\, ,\, \Phi) \neq 0 \ , 
\end{equation}
where 
\begin{equation}\label{eq:mukai}
	(A,B)\equiv (A\wedge \lambda(B))_{10}\ 
\end{equation}
is the Chevalley--Mukai pairing (${}_{10}$ denotes keeping the ten-form part only). The reason to demand (\ref{eq:comp}) is that, for both $i=1,2$, we have $\overline{\epsilon_i} \gamma_{+_i} \epsilon_i= 32 K^i e_{+_i}= 32 g_{-_i +_i}=16$, which is non-zero. (Once we determine a metric, (\ref{eq:comp}) will be equal to $16^2 {\rm vol}_{10}$.)

To see whether adding the data (\ref{eq:+1+2}) to $\Phi$ allows us to determine a globally defined metric, we have to compute the common stabilizer of $\Phi$ and of (\ref{eq:+1+2}). An element $\omega_{AB}\Gamma^{AB}$ of the Lie algebra so(10,10) acts on $T \oplus T^*$ as a commutator:
\begin{equation}
	[\omega_{AB} \Gamma^{AB}, \ \cdot\ ]\ .
\end{equation}
We see now that all of the offending elements in ${\rm stab}(\Phi)$, of the form $\stackrel \to \gamma_M \stackrel \leftarrow \gamma_N$, contain either a $\stackrel\to \gamma_{-_1}$ or a $\stackrel \leftarrow \gamma_{-_2}$, and so they do not commute with one or both of the two elements (\ref{eq:+1+2}). The resulting stabilizer is
\begin{equation}\label{eq:stabPhi+1+2}
	{\rm stab}(\Phi, \stackrel\to \gamma_{+_1},\stackrel \leftarrow \gamma_{+_2})={\rm span}\{\omega_{\bf 21}^{\alpha_1 \beta_1} \stackrel\to\gamma_{\alpha_1 \beta_1}\, ,\ \omega_{\bf 21}^{\alpha_2 \beta_2} \stackrel\leftarrow\gamma_{\alpha_2 \beta_2}\}\ .
\end{equation}
This is now contained in the Lie algebra so(10,10) (it is in fact simply ${\rm spin}(7)\oplus {\rm spin}(7)$). So the three data 
\begin{equation}
	(\Phi, \stackrel\to \gamma_{+_1},\stackrel \leftarrow \gamma_{+_2})
\end{equation}
do determine a globally defined metric and $B$ field. 

Notice, however, that we have been overzealous in adding $\stackrel\to\gamma_{+_1}$ and $\stackrel \leftarrow \gamma_{+_2}$ to $\Phi$: in addition to the elements outside ${\rm O}(9,1)\times {\rm O}(9,1)$, we have also fixed $\stackrel \to \gamma_{\alpha_1 -_1}$ and $\stackrel \leftarrow \gamma_{\alpha_2 -_2}$, which are in the Lie algebra of ${\rm O}(9,1)\times {\rm O}(9,1)$. In fact, the stabilizer we would have wanted to obtain is 
\begin{equation}
	{\rm stab}(g_{MN},B_{MN},\epsilon_1,\epsilon_2)= {\rm span}
	\{\omega_{\bf 21}^{\alpha_1 \beta_1} \stackrel\to\gamma_{\alpha_1 \beta_1}\, ,\ \stackrel \to \gamma_{-_1 \alpha_1}\, , \ \omega_{\bf 21}^{\alpha_2 \beta_2} \stackrel\leftarrow\gamma_{\alpha_2 \beta_2}\, , \ \stackrel\leftarrow \gamma_{-_2 \alpha_2}\}\ ,
\end{equation}
which has the Lie algebra structure of the group ${\rm ISpin}(7)\times {\rm ISpin}(7)$. This means that we are parameterizing $(g,B,\epsilon_1,\epsilon_2)$, but that part of the information in $\stackrel\to\gamma_{+_1}$ and $\stackrel \leftarrow \gamma_{+_2}$ is spurious. This is not a big issue for our purposes: it gives rise to a potential topological subtlety in using a ${\rm Spin}(7)\times {\rm Spin}(7)$ structure rather than an ${\rm ISpin}(7)\times {\rm ISpin}(7)$ structure, but this should be of very little importance. Moreover, these spurious data are quantifiably few, as we will see shortly. 

Since ${\rm Spin}(7)\times {\rm Spin}(7)$ is a subgroup of ${\rm O}(8)\times {\rm O}(8)$ and not just of ${\rm O}(9,1)\times {\rm O}(9,1)$, our data now potentially give rise to more than one globally defined metric (see footnote \ref{foot:id}). So we should specify the procedure to obtain the metric and $B$ field. Concretely, one can extract $E_{MN}\equiv g_{MN}+B_{MN}$ as follows. First of all, we can identify $\stackrel \to \gamma_{-_1}$ as the element of ${\rm Ann}(\Phi)$ which anticommutes with $\stackrel \leftarrow \gamma_{+_2}(-)^{\rm deg}$. From the stabilizer (\ref{eq:stabPhi+1+2}) we can now also find eight more elements $\stackrel \to \gamma_{\alpha_1}$. This gives us a basis for $C_+$, the bundle of left-acting gamma matrices. Now we can extract $E_{MN}$ from (\ref{eq:gammaE}). (We can run the same procedure with the right-acting gamma matrices $\stackrel \leftarrow \gamma_M$, with the same results.) 

Now that we have determined a choice of $g$ and $B$, we can write any bispinor as $e^B\sum_{i=1}^{32} \epsilon_{1i}\otimes \overline{\epsilon_{2i}}$ in terms of some spinors $\epsilon_{1i}$ and $\epsilon_{2i}$. But, if more than one of the $\epsilon_{1i}$ were non-zero, the elements of the type $\stackrel \to \gamma_{MN}$ in (\ref{eq:stabPhi+1+2}) would not have the Lie algebra structure of ${\rm Spin}(7)$, but of a subgroup. So we see that only one of the $\epsilon_{1i}$ should be non-vanishing; and similarly, only one of the $\epsilon_{2i}$. So we can write $\Phi=e^B \epsilon_1 \otimes \overline{\epsilon_2}$ for some $\epsilon_{1,2}$. Moreover, we can determine these two spinors from their stabilizers. In the basis introduced in section \ref{ssub:structure}, both $\epsilon_i$ can be written as $| \!\uparrow\, \rangle  \otimes \eta_{8,i} $; both $\eta_{8,i}$ can then be determined from the elements $\stackrel \to \gamma_{\alpha_1 \beta_1}$, $\stackrel \leftarrow \gamma_{\alpha_2 \beta_2}$ in ${\rm stab}(\Phi)$. 

So we have seen that from $(\Phi, \stackrel\to \gamma_{+_1},\stackrel \leftarrow \gamma_{+_2})$ we can reconstruct $(g,B,\epsilon_{1,2})$. Since $\Phi$ can be written as $e^B\epsilon_1 \otimes \overline{\epsilon_2}$, and since the action of Spin(9,1) on the space of Weyl spinors is transitive, we see that the action of so(10,10) on the space of generalized ISpin(7) structures is also transitive. We can then compute the dimension of this space as the dimension of so(10,10) minus the dimension of ${\rm stab}(\Phi)$ in (\ref{eq:stabPhi}): this gives 190-78=112. We see that this is 20 less than the degrees of freedom of the data $(g,B,\epsilon_{1,2})$, which are $10^2+2\times 16=132$. So we see once again that $\Phi$ does not contain enough data.

On the other hand, $\stackrel\to\gamma_{+_1}$ and $\stackrel \leftarrow \gamma_{+_2}$ are two elements of $T \oplus T^*$, so each has 20 degrees of freedom. Together with the 112 degrees of freedom of $\Phi$, this brings us to 152, which is 20 more than we needed to match $(g,B,\epsilon_{1,2})$. This quantifies the redundancy in our parameterization $(\Phi, \stackrel\to \gamma_{+_1},\stackrel \leftarrow \gamma_{+_2})$.


\subsection{Summary of this section} 
\label{sub:sumgen}

We have shown that the degrees of freedom $(g_{MN}, B_{MN}, \epsilon_1, \epsilon_2)$ can be reformulated in terms of the data $(\Phi, \stackrel\to \gamma_{+_1}, \stackrel \leftarrow \gamma_{+_2})$, where:
\begin{itemize}
	\item $\Phi$ is a ``generalized ISpin(7) structure'', namely a form whose stabilizer in Spin(10,10) is the group (\ref{eq:genISpin}); practically speaking, this just means that at every point $\Phi$ can be written in one of the ways listed in (\ref{eq:PhiIIA}) or (\ref{eq:PhiIIB}). $\Phi$ has a two-dimensional annihilator ${\rm Ann}(\Phi)\subset T \oplus T^*$. This can be thought of as generated by left Clifford action by a null vector $K_1$, and right action by a vector $K_2$. But $\Phi$ alone does not determine a metric. Simplifying quite a bit our discussion in section \ref{sub:metric}, we can say that it gives nine elements $\{e_{-_1}=K_1,e_{\alpha_1}\}_{\alpha_1=1,\ldots,8}$ of a ``left'' vielbein, and nine elements $\{e_{-_2}=K_2,e_{\alpha_2}\}_{\alpha_2=1,\ldots,8}$ of a ``right'' vielbein; both of these vielbeine are incomplete.
	\item $\stackrel\to \gamma_{+_1}= e_{+_1}\cdot (\ )$ and $\stackrel \leftarrow \gamma_{+_2}= (\ )\cdot e_{+_2}$ are two elements of $T \oplus T^*$. We can think of $e_{+_1}$ and $e_{+_2}$ as completing the left and right vielbeine mentioned earlier.
\end{itemize}
The fact that the $e_{+_i}$ represent the missing vectors in the vielbeine defined by $\Phi$ is enconded in the ``compatibility condition'' (\ref{eq:comp}) among the data $(\Phi, \stackrel\to \gamma_{+_1}, \stackrel \leftarrow \gamma_{+_2})$. 

Within ${\rm Ann}(\Phi)$, a distinctive role will be played by the element that commutes with $\stackrel \to \gamma_{+_1} - \stackrel \leftarrow \gamma_{+_2}(-)^{\rm deg}$, namely $\stackrel \to \gamma_{-_1} - \stackrel \leftarrow \gamma_{-_2}(-)^{\rm deg}$, which we called $\tilde K \wedge + K \llcorner$ (see (\ref{eq:KKtPhi})). 

\bigskip

In the next section, we will reformulate the conditions for unbroken supersymmetry on $(g_{MN}, B_{MN}, \epsilon_1, \epsilon_2)$ as differential equations on $(\Phi, \stackrel\to \gamma_{+_1}, \stackrel \leftarrow \gamma_{+_2})$.



\newpage
\section{Differential equations} 
\label{sec:diff}

We will now see how the supersymmetry conditions look like in terms of the forms summarized in section \ref{sub:sumgen}. 
 
\subsection{Necessary and sufficient system} 
\label{sub:diff}

We give here the differential equations, relegating their derivation to appendix \ref{app:suff}. In both IIA and IIB\footnote{To make (\ref{eq:psp10}) identical in IIA and IIB, we have changed the conventions for IIA with respect to \cite{democratic} by setting $F_{\rm here}= \lambda(F_{\rm there})$, and $H_{\rm here}=-H_{\rm there}$. (This differs by a sign from the redefinition used in \cite{gmpt3}.) We still have a sign of difference among the two theories: the upper sign in (\ref{eq:++1}) is for IIA, the lower for IIB.}, they read
\begin{subequations}\label{eq:susy10}
	\begin{align}
		\label{eq:psp10}
		&\framebox{$d_H(e^{-\phi} \Phi) = -(\tilde K\wedge + \iota_K ) F$} \ ;\\
		\label{eq:LK}&L_K g = 0 \ ,\qquad d\tilde K = \iota_K H \ ;\\
		& \label{eq:++1} 
		\left(e_{+_1}\cdot \Phi \cdot e_{+_2}\, ,\, \gamma^{MN} \left[ \pm d_H(e^{-\phi} \Phi \cdot e_{+_2}) + \frac12 e^\phi d^\dagger (e^{-2 \phi} e_{+_2})\Phi - F\right]\right)=0\ ;\\
		& \label{eq:++2}
		\left(e_{+_1}\cdot \Phi \cdot e_{+_2}\, ,\, \left[ d_H(e^{-\phi} e_{+_1}\cdot\Phi) - \frac12 e^\phi d^\dagger (e^{-2 \phi} e_{+_1})\Phi - F\right] \gamma^{MN}\right)=0\ .
	\end{align}
\end{subequations}
Here, $g$ is the metric, $\phi$ is the dilaton, $H$ is the NSNS three-form, $d_H\equiv d-H\wedge$, and $F$ is the ``total'' RR field strength $F=\sum F_k$. In the spirit of the ``democratic'' approach championed in \cite{democratic}, the sum is from $0$ to $10$ in IIA and from 1 to 9 in IIB, and one cuts the number of forms down by half with the self-duality constraint 
\begin{equation}\label{eq:F*F}
	F=* \lambda (F) \ .
\end{equation}
Finally, $(\ , \ )$ is the usual pairing on forms given in (\ref{eq:mukai}).

Equations (\ref{eq:susy10}) are \emph{necessary and sufficient} for supersymmetry to hold; we give some details of this computation in appendix \ref{app:suff}. To also solve the equations of motion, one needs to impose the Bianchi identities, which away from sources (branes and orientifolds) read 
\begin{equation}\label{eq:bianchi}
	dH=0 \ ,\qquad d_H F=0\ .
\end{equation}
It is then known (see \cite{lust-tsimpis} for IIA, \cite{gauntlett-martelli-sparks-waldram-ads5-IIB} for IIB) that almost all of the equations of motion for the metric and dilaton follow. 

The system (\ref{eq:susy10}) is equivalent to the minimum amount of supersymmetry: one supercharge only. To have more than one supercharge, we should simply demand that there be several solutions to (\ref{eq:susy10}), which share the same the physical fields (the metric, dilaton, and fluxes). An example of this will actually be seen in section \ref{sub:mink4}, where we will apply~(\ref{eq:susy10}) to four-dimensional vacuum solutions, which have at least four supercharges. In that case, we will indeed find four independent solutions to (\ref{eq:susy10}).

(\ref{eq:psp10}) is similar to an equation found to be necessary for supersymmetry in \cite{koerber-martucci-ads} in a 1+9 splitting. It is one of the main result in this paper: as we will see later, this equation by itself reproduces all of the ``pure spinor'' equations for four-dimensional vacua \cite{gmpt2}. The next two equations, (\ref{eq:LK}), already appeared in \cite{hackettjones-smith,figueroaofarrill-hackettjones-moutsopoulos,koerber-martucci-ads}. Unfortunately, the other main result of our paper is that (\ref{eq:psp10}) and (\ref{eq:LK}), albeit necessary for supersymmetry, are not sufficient. Schematically, the problem with (\ref{eq:psp10}) is that the covariant derivatives 
\begin{equation}\label{eq:nabla+}
	(\nabla_{+_2}+ H_{+_2}) \epsilon_1 \ ,\qquad (\nabla_{+_1}- H_{+_1}) \epsilon_2 
\end{equation}
are completely absent from $d \Phi$. (\ref{eq:LK}) do contain some components of (\ref{eq:nabla+}), but not all of them. So one has to look for a way of re-expressing the missing components in terms of differential forms; in appendix \ref{app:suff}, we show that (\ref{eq:++1}) and (\ref{eq:++2}) do the job. They are perhaps not as nice as one might have wished, because they do contain the metric explicitly in the Clifford products. On a more positive note, they do not contain any covariant derivative. Moreover, the system (\ref{eq:susy10}) is necessary and sufficient for both IIA and IIB, and for any $\Phi$ among the many possibilities we gave in (\ref{eq:PhiIIA}), (\ref{eq:PhiIIB}), or interpolating among those. It is of course possible that a better version will be found in the future. In any case, for the vast majority of situations (\ref{eq:psp10}) and (\ref{eq:LK}) will actually be enough, and~(\ref{eq:++1}),~(\ref{eq:++2}) will not contain any new information. For example, we will see in section \ref{sub:mink4} that this is the case for four-dimensional vacua. This is because in that situation (\ref{eq:nabla+}) are related by four-dimensional Lorentz symmetry to other components $\nabla_\mu \epsilon$, whose equations are already implied by (\ref{eq:psp10}) and (\ref{eq:LK}). On the other hand, sometimes (\ref{eq:++1}) and (\ref{eq:++2}) do contain some new information, as we will see in section \ref{sub:mink3}, where they will be seen to reproduce a constraint first noticed in \cite{haack-lust-martucci-t,smyth-vaula}. 


\subsection{Symmetry} 
\label{sub:symm}

The first equation in (\ref{eq:LK}) tells us that $K$ is an isometry. Notice that the second in~(\ref{eq:LK}) also implies $L_K H=0$. We also show at the end of section \ref{sub:psp10QT} that (\ref{eq:psp10}), (\ref{eq:LK}) imply $L_K\phi=0$.

We will now also show that $L_K F=0$. Notice first that 
\begin{equation}\label{eq:LKcomm}
	\{ d_H, \tilde K \wedge + \iota_K \} = \{ d, \tilde K\wedge \} - \{ H\wedge, \iota_K\} + \{d , \iota_K\} = (d \tilde K - \iota_K H)\wedge + L_K= L_K\ ,
\end{equation}
where, in the last step, we have used (\ref{eq:LK}). 
Now, if one acts on (\ref{eq:psp10}) with $d_H$, the left hand side vanishes since $d_H^2=0$. The right hand side then gives
\begin{equation}
	\begin{split}
		0= d_H ((\tilde K \wedge + \iota_K ) F)=&\{ d_H, (\tilde K\wedge + \iota_K )\} F - (\tilde K\wedge + \iota_K ) d_H F\\
		&=L_K F - (\tilde K + \iota_K ) d_H F\ ;
	\end{split}
\end{equation}
so if one imposes the Bianchi identities (\ref{eq:bianchi}), one gets 
\begin{equation}
	L_K F=0\ ,
\end{equation}
as promised. So we see that $K$ is not just an isometry, but a symmetry of the full solution \cite{figueroaofarrill-hackettjones-moutsopoulos}.

In fact, $K$ is also a supersymmetric isometry: namely, it keeps $\Phi$ invariant. To see this, act on (\ref{eq:psp10}) with $\tilde K \wedge + \iota_K$. The left-hand side gives,
\begin{equation}
	(\tilde K \wedge + \iota_K) d_H(e^{-\phi} \Phi) = L_K (e^{-\phi} \Phi) - d_H (e^{-\phi}(\tilde K \wedge + \iota_K) \Phi)= L_K(e^{-\phi}\Phi) \ ;
\end{equation}
we have used (\ref{eq:LKcomm}), and (\ref{eq:KKtPhi}). The right-hand side gives 
\begin{equation}
	-\frac12(\tilde K \wedge + \iota_K)^2 F = - \tilde K \cdot K F = -\frac14 (K_1^2 - K_2^2) F = 0 \ ,
\end{equation}
where we have used (\ref{eq:KKt}) and that $K_{1,2}$ are null (from (\ref{eq:Knull}) applied to both spinors $\epsilon^{1,2}$). Recalling from (\ref{eq:LKphi}) that $L_K \phi=0$, we conclude
\begin{equation}
	L_K \Phi = 0 \ .
\end{equation}


\section{Examples} 
\label{sec:examples}

We will now see how (\ref{eq:susy10}) reproduces known equations in particular setups. 

\subsection{Minkowski$_4$} 
\label{sub:mink4}

We will start by considering four-dimensional Minkowski vacua, namely solutions of the form $\rr^{3,1}\times M_6$ where all fields preserve the maximal symmetry of the four-dimensional space. We will recover the equations considered in \cite{gmpt2}. As we will see, they all come from (\ref{eq:psp10}).

\subsubsection{Structure of four-dimensional spinors} 
\label{ssub:4dstructure}
We first need to understand the geometry of four-dimensional spinors, similarly to our discussion in section \ref{sec:gen} of Majorana--Weyl spinors in ten dimensions. As is well-known, in four dimensions one can impose either the Majorana or Weyl conditions, but not both. Moreover, one can map a Majorana spinor to a Weyl one, and viceversa. We will consider a Weyl spinor $\zeta_+$ of positive chirality. We will also define its conjugate $\zeta_- = (\zeta_+)^*$, which has negative chirality. (We will work in the basis where all $\gamma_\mu$ are real.) There are two bispinors we can consider: $\zeta_+ \otimes \overline{\zeta_+}$ and $\zeta_+ \otimes \overline{\zeta_-}$. We will investigate them in turn. 

We start by looking at $\zeta_+ \otimes \overline{\zeta_+}$, using the Fierz identity (\ref{eq:fierzd}) for $d=4$. Since $\zeta_+$ is chiral, we see that only bilinears with an odd number of $\gamma_\mu$'s survive. That leaves us with
\begin{equation}\label{eq:z+z+}
	\zeta_+ \otimes \overline{\zeta_+} = (1+\gamma)v = v + i *v\ ,
\end{equation}
where $v^\mu \equiv \frac14 \overline{\zeta_+} \gamma^\mu \zeta_+$ is a real vector. A similar computation as (\ref{eq:Ke}) also reveals that
\begin{equation}\label{eq:vz}
	v \,\zeta_+ = 0 \ .
\end{equation}
This also implies that $v$ is null: 
\begin{equation}
	v^\mu v_\mu =0 \ .
\end{equation}

Moving on to $\zeta_+ \otimes \overline{\zeta_-}$, by chirality only bilinears with an even number of $\gamma_\mu$'s survive in the Fierz identity (\ref{eq:fierzd}). Moreover, the bilinear $\overline{\zeta_-} \gamma^{\mu_1 \ldots \mu_k} \zeta_+= \zeta_+^t \gamma^0 \gamma^{\mu_1 \ldots \mu_k} \zeta_+$ vanishes when $\gamma^0 \gamma^{\mu_1\ldots \mu_k}$ is antisymmetric. That leaves us with only $k=2$. So 
\begin{equation}
	\zeta_+ \otimes \overline{\zeta_-}= \omega_+\ ;
\end{equation}
from (\ref{eq:stard}) (with $d=4$; we pick the convention $c=i$) and the fact that $\zeta_+$ is chiral we see that $\omega_+ = -i * \omega_+$. Also, (\ref{eq:vz}) and (\ref{eq:gwedge}) tells us
\begin{equation}
	v\wedge \omega_+ = \iota_v \omega_+ = 0 \ ,
\end{equation}
so in fact $\omega_+= v \wedge w$ for some complex one-form $w$, which also satisfies $\iota_v w=0$ (recall that $v$ is null, and that $w$ is complex). So, summing up:
\begin{equation}\label{eq:vw}
	\zeta_+ \otimes \overline{\zeta_-} = v \wedge w\ ,
\end{equation}
with $v$ the same as in (\ref{eq:z+z+}), and $v^\mu v_\mu = v^\mu w_\mu=0$.

Although we will not quite need it in what follows, we can also now compute the structure group of $\zeta_+$. We can compute it geometrically as the stabilizer of $v$ and $w$ under the Lorentz action, similarly to section \ref{ssub:structureforms}. Since $v$ is null, its stabilizer is ${\rm ISO}(2)= {\rm SO}(2)\ltimes \rr^2$. As for $w$, we see that it plays the role of $\Psi_4$ in section \ref{ssub:structureforms} (compare (\ref{eq:O4}) and (\ref{eq:vw})). 
In ten dimensions, $\Psi_4$ breaks from SO(8) to Spin(7); in four dimensions, $w$ breaks SO(2) completely. We are thus left with a structure group isomorphic to $\rr^2$. 

This can actually be seen directly from the spinor perspective. If one uses that ${\rm Spin}(3,1)\cong {\rm Sl}(2,\cc)$, one can see that the stabilizer of e.g.~the spinor ${1 \choose 0}$ is the set of matrices of the form ${{1 \ z}\choose {0 \ 1}}$, which is isomorphic to $\cc\cong \rr^2$.

Finally, using the same logic as in (\ref{eq:hopf}), one can also see that the map $\zeta \mapsto v$ is again a Hopf fibration; this time it reproduces the classic $S^3 \to S^2$ with fibre $S^1$.
 

\subsubsection{Reproducing the pure spinor equations for four-dimensional vacua} 
\label{ssub:mink4repr}

For a four-dimensional vacuum solution, Poincar\'e invariance fixes the metric to be 
\begin{equation}\label{eq:metric46}
	ds^2_{10}= e^{2A} ds^2_4 + ds^2_6 \ .
\end{equation}
Moreover, the flux $F$ should be
\begin{equation}\label{eq:Ff}
	F= f + e^{4A} {\rm vol}_4 \wedge *_6 \lambda f\ ,
\end{equation}
for $f$ a form on the internal manifold $M_6$. $H$ is constrained to be purely internal.

We now proceed to splitting the ten-dimensional spinors $\epsilon_{1,2}$ in terms of the four-dimensional spinors we just studied. For an ${\cal N}=1$ four-dimensional vacuum, this decomposition reads
\begin{equation}\label{eq:e46}
	\begin{split}
		&\epsilon_1 = \zeta_+  \eta^1_+ + \zeta_- \eta^1_-\ ,\\
		&\epsilon_2 = \zeta_+  \eta^2_\mp + \zeta_- \eta^2_\mp\ ;
	\end{split}
\end{equation}
here and later, the upper sign is for IIA, the lower for IIB. By definition, $\eta^a_-=(\eta^a_+)^*$, so that both $\epsilon_{1,2}$ are Majorana--Weyl. Poincar\'e invariance demands that the $\epsilon_i$ in (\ref{eq:e46}) be supersymmetric for any $\zeta_+$; this will give a total of four supercharges, as appropriate for an ${\cal N}=1$ vacuum. 
Using this decomposition, we can now specialize the ingredients in (\ref{eq:psp10}): namely the bispinor $\epsilon$ and its annihilators $K_\pm$. The ten-dimensional gamma matrices can be decomposed as 
\begin{equation}\label{eq:gamma46}
	\gamma^{(10)}_\mu = e^A\gamma^{(4)}_\mu\otimes 1 \ ,\qquad
	\gamma^{(10)}_m = \gamma_5 \otimes \gamma^{(6)}_\mu \ ,
\end{equation}
where $\{\gamma^{(d)}\}$ is a basis of gamma matrices in $d$ dimensions, and $\gamma_5$ is the chiral operator in four dimensions.

The bispinor $\Phi= \epsilon_1 \otimes \overline{\epsilon_2}$ can be evaluated using the ten-dimensional Fierz identities~(\ref{eq:ee}), and repackaged using the ones (\ref{eq:fierzd}) in four and six dimensions; we get 
\begin{equation}\label{eq:Phi46}
\begin{split}
		\Phi&= \mp(\zeta_+ \overline{\zeta_+})\wedge (\eta^1_+ \eta^{2\,\dagger}_\mp) + (\zeta_- \overline{\zeta_-}) \wedge (\eta^1_- \eta^{2\,\dagger}_\pm) + (\zeta_+ \overline{\zeta_-}) \wedge (\eta^1_+ \eta^{2\,\dagger}_\pm) \pm (\zeta_- \overline{\zeta_+}) \wedge (\eta^1_- \eta^{2\,\dagger}_\mp)\\
		&=\mp 2 {\rm Re} ((e^A v + i e^{3A}*_4 v)\wedge \phi_\mp)
		+ 2 {\rm Re} ( e^{2A} v\wedge w \wedge \phi_\pm)\ .
\end{split}
\end{equation}
$\phi_\pm = \eta^1_+ \otimes \eta^{2\,\dagger}_\pm$ are two six-dimensional pure spinors associated to the internal geometry; they define together an ${\rm SU}(3) \times {\rm SU}(3)$ structure.

We can similarly evaluate $K_1$ and $K_2$. First, recall that for a six-dimensional Weyl spinor $\eta_+$ we have $\eta_-^\dagger \gamma^m \eta_+=0$, with $m=1,\ldots, 6$ being an internal index. So $K_i^m=0$. For the spacetime components, from the definition (\ref{eq:KM}) and from the spinorial decomposition (\ref{eq:e46}) we get 
\begin{equation}\label{eq:Keta2}
	K_i^\mu = \frac1{32}e^{-A}\left(\overline{\zeta_+} \gamma^\mu \zeta_+ || \eta^i_+ ||^2 + \overline{\zeta_-} \gamma^\mu \zeta_- || \eta^i_- ||^2\right)= \frac18 e^{-A} v^\mu || \eta^i ||^2
\end{equation}
Let us assume for simplicity $|| \eta^1_+ ||=|| \eta^2_+ ||\equiv || \eta ||^2$. Then we have $\tilde K=0$, $K=K_1=K_2$ (recall (\ref{eq:KKt})). 

Now we can start applying our differential equations (\ref{eq:susy10}). We start from (\ref{eq:LK}). This tells us that $K=\frac18 v (e^{-A} || \eta ||^2)$ should be a Killing vector. But by construction $v$ is already a Killing vector; so we get
\begin{equation}
	|| \eta ||^2 = c_+ e^A
\end{equation}
for some constant $c_+$. This was indeed the condition on the spinor norm found in \cite{gmpt2} (in the present simplifying assumption that $|| \eta_1 ||^2 = ||  \eta_2 ||^2$). To summarize so far, we have
\begin{equation}\label{eq:KKt46}
	\tilde K = 0 \ ,\qquad K= \frac18 c_+ v\ .
\end{equation}

Now we can turn to (\ref{eq:psp10}). We have already evaluated $\Phi$ in (\ref{eq:Phi46}). So the left-hand side of (\ref{eq:psp10}) is easy to evaluate: since everything is indepedent of four-dimensional spacetime, $d_H$ only acts on $M_6$. As for the right-hand side of (\ref{eq:psp10}), using (\ref{eq:Ff}) and (\ref{eq:KKt46}):
\begin{equation}\label{eq:KF4}
	- (\tilde K \wedge + \iota_K) F = - \frac18 c_+ \iota_v (e^{4A} {\rm vol_4}\wedge *_6 \lambda f)= \frac18 c_+ *_4 v \wedge e^{4A} *_6 \lambda	 f \ .
\end{equation}
We can now use (\ref{eq:KF4}) and (\ref{eq:Phi46}) in (\ref{eq:psp10}); we get
\begin{subequations}\label{eq:psp46}
	\begin{align}
		&d_H {\rm Re} (e^{A-\phi} \phi_\mp) = 0 \ ,\qquad
		\\
		&d_H (e^{2A-\phi} \phi_\pm) = 0 \ , \label{eq:psp46closed}
		\\
				&d_H {\rm Im} (e^{3A-\phi} \phi_\mp) = \mp\frac1{16}c_+ e^{4A} *_6 \lambda \, f \ .\qquad \label{eq:psp46f}
	\end{align}	
\end{subequations}
These are the pure spinor equations found in \cite{gmpt2} (in the simplifying assumption $|| \eta_1 ||^2 = || \eta_2 ||^2$); the upper sign is for IIA, the lower for IIB. So we see that (\ref{eq:psp10}) reproduces them all in one go. Notice also that nothing in our computation depended on a particular choice of $v$ and $w$; this means that we have found not just one solution to (\ref{eq:psp10}), but four --- as many as the number of Weyl spinors $\zeta_+$; this corresponds to four supercharges, as appropriate for an ${\cal N}=1$ solution.

We still have to look at (\ref{eq:++1}), (\ref{eq:++2}). We do not expect anything new from these, since we have already reproduced all the equations found in \cite{gmpt2}. We will look at (\ref{eq:++1}); the analysis of (\ref{eq:++2}) is similar. First, we have to choose $e_{+_1}\cdot$ and $\cdot e_{+_2}$. Since $K_1=K_2=K$, it also makes sense to take $e_{+_1}=e_{+_2}\equiv e_+$. Moreover, since $K\propto v$ has only four-dimensional components, we will take this to be the case for $e_+$ as well. More specifically, to take care of the powers of the warping $e^A$ coming from (\ref{eq:gamma46}), we will write
\begin{equation}\label{eq:gamma+46}
	\stackrel \to \gamma_+  = e^A e_+ \wedge + e^{-A} e_+ \llcorner \ ,\qquad
	\stackrel \leftarrow \gamma_+(-)^{\rm deg} = e^A e_+ \wedge - e^{-A} e_+ \llcorner\ ,
\end{equation}
where now $\llcorner$ refers to a contraction using the four-dimensional metric. Since $e_+$ has no internal components, $d^\dagger(e^{-2 \phi}e_+)$ vanish. Also, from (\ref{eq:gamma+46}) we get
\begin{equation}\label{eq:dgamma+46}
	\{ d, \stackrel \leftarrow \gamma_+ (-)^{\rm deg} \}=
	\{ dx^M\wedge \del_M ,  e^A e_+ \wedge - e^{-A} e_+ \llcorner\}=
	e^{-A} \del_+ + dA\wedge \stackrel \to \gamma_+\ .
\end{equation}
We now apply all this to (\ref{eq:++1}):
\begin{equation}\label{eq:++146}
	(\gamma_+ \Phi \gamma_+ , \gamma^{MN}\left[ \pm d_H(e^{-\phi} \Phi \gamma_+ )-F\right])= (\gamma_+ \Phi \gamma_+, \gamma^{MN}[dA\wedge \gamma_+ e^{-\phi} \Phi-2f])\ ;
\end{equation}
we have used (\ref{eq:dgamma+46}) and (\ref{eq:Ff}). Let us now consider the various possibilities for the indices $MN$. If $M=m$ and $N=n$, both terms in (\ref{eq:++146}) vanish because $\gamma_+^2=0$, and the equation has no content. For $M=\mu$, $N=\nu$, the $dA \wedge \gamma_+ \Phi$ term drops out, and we are left with
\begin{equation}\label{eq:++146munu}
	(\gamma_+ \Phi \gamma_+, \gamma^{\mu\nu} \, f)= \mp \frac1{32}\overline{\epsilon_1} \gamma_+ \gamma^{\mu\nu}f \gamma_+ \epsilon_2 = -\frac1{32}(\overline{\zeta_+} \gamma_+ \gamma^{\mu\nu} \gamma_+ \zeta_-) (\eta_+^{1\,\dagger}f \eta_\pm^2+{\rm c.~c.})\ ;
\end{equation}
we have used (\ref{eq:egCge}) and the spinor decomposition (\ref{eq:e46}). This implies ${\rm Tr}(\phi_\pm^\dagger f)=0$; or, in terms of the six-dimensional Chevalley--Mukai pairing,
\begin{equation}\label{eq:*fphi}
	(\lambda *f,\phi_\pm)=0\ .
\end{equation}
This actually follows from (\ref{eq:psp46f}) and from $(d \phi_\pm, \phi_\mp)= ( \phi_\pm, d \phi_\mp)$. 

Finally, for $M=m$ and $N=\nu$, a computation similar to (\ref{eq:++146munu}) gives 
\begin{equation}\label{eq:conseq}
	\left(dA \wedge \phi_\pm \pm i \frac{c_+}{16} e^{3A} * \lambda f, \gamma^m  \overline{\phi_\pm}\right)=0\ .
\end{equation}
This can also be shown to follow from (\ref{eq:psp46}). Using the ``intrinsic torsions'' introduced for example in \cite[Eq.~(A.19)]{gmpt3} and the expression of the pure spinors as a bispinor, $\phi_\pm = \eta^1_+ \otimes \eta^{2\,\dagger}_\pm$, one derives that in generalized complex geometry $(d \phi_\mp, \gamma^m \phi_\mp)$; computing the $d \phi_\mp$ from (\ref{eq:psp46f}), one then obtains (\ref{eq:conseq}).

In conclusion, for four-dimensional Minkowski vacua, the system (\ref{eq:susy10}) reproduces the conditions found in \cite{gmpt2}. Equation (\ref{eq:LK}) reduces to a condition about the norm of the spinors, while (\ref{eq:psp10}) reproduces all of the pure spinor equations in \cite{gmpt2}. (\ref{eq:++1}) and (\ref{eq:++2}) are, in this case, redundant. 



\subsection{Minkowski$_3$} 
\label{sub:mink3}

We will now consider solutions of the form $\rr^{2,1}\times M_7$. One application of these vacua is to the study three-dimensional RG flows holographically. In the context of this paper, this case will be an example where equations (\ref{eq:++1}), (\ref{eq:++2}) are not redundant; they will reproduce an algebraic constraint found in \cite{haack-lust-martucci-t,smyth-vaula}.  
 
Most of our discussion is similar to the one in section \ref{sub:mink4}, so we will be more schematic here.   

We will first discuss the forms of three-dimensional spinors. This is much simpler than in section \ref{ssub:4dstructure}.  We can work in a basis where all the gamma matrices are real (and symmetric). It then follows that the ``norm'' $\overline{\zeta}\zeta =0$ for any spinor  $\zeta$ (just as we found in ten dimensions). Applying (\ref{eq:fierzodd}) in either the even or the odd version we find
\begin{equation}\label{eq:zz3}
	\zeta \otimes \overline{\zeta} = v = -*_3\,v\ ,
\end{equation}
where as usual we are omitting the slash: $v= \rlap{\begin{picture}(10,10)(2,2)
\put(0,0){\line(1,1){10}}
\end{picture} }v = v_M \gamma^M$, and similarly for $*_3v$.

We now need to decompose our ten-dimensional spinor in terms of three- and seven-dimensional ones. The decomposition of gamma matrices is a bit harder than (\ref{eq:gamma46}):
\begin{equation}
	\gamma^{(10)}_\mu = e^A\sigma_3 \otimes \gamma^{(3)}_\mu \otimes 1
	\ ,\qquad
	\gamma^{(10)}_m = \sigma_1 \otimes 1 \otimes \gamma^{(7)}_m\ ,
\end{equation}
where $\sigma_i$ are Pauli matrices. Taking $\gamma^{(3)}_\mu$ all real and $\gamma^{(7)}_m$ purely imaginary, all the $\gamma^{(10)}_M$ are real. Our two ten-dimensional Majorana--Weyl spinor now can be decomposed as
\begin{equation}\label{eq:e37}
	\epsilon_1 = {1\choose -i}\otimes \zeta \otimes \eta^1
	\ ,\qquad
	\epsilon_2 = {1\choose i}\otimes \zeta \otimes \eta^2\ ,
\end{equation}
where $\zeta$, $\eta^1$, $\eta^2$ are all real. We can use the Fierz identities (\ref{eq:ee}) in ten dimensions, together with the ones (\ref{eq:fierzodd}) in 3 and 7 dimensions, to obtain
\begin{equation}
	\begin{split}
		\Phi&=(\zeta\otimes \overline{\zeta})_+\wedge \phi_\pm \mp 
		(\zeta\otimes \overline{\zeta})_- \wedge\phi_\mp=\\
		&=-*_3 v \wedge\phi_\pm + v\wedge \phi_\mp
		\ ,
	\end{split}
\end{equation}
where $\phi= \eta\otimes \eta^\dagger$. The subscript $()_\pm$ refers to the possibility of obtaining an even or odd form in (\ref{eq:fierzodd}); $(\zeta\otimes \overline{\zeta})_\pm$ have then been evaluated using (\ref{eq:zz3}). As for $K$ and $\tilde K$, more or less the same computation that led us to (\ref{eq:KKt46}) now gives us
\begin{equation}
	\tilde K = 0 \ ,\qquad K = c_+ v\ ,
\end{equation}
again under the simplifying assumption $|| \eta_1 ||^2=|| \eta_2 ||^2$.

We can now apply (\ref{eq:psp10}), using the same steps as in section \ref{ssub:mink4repr}. We get
\begin{subequations}\label{eq:psp37}
	\begin{align}
			&d_H(e^{A-\phi} \phi_\mp)=0\ ,\\
			&d_H(e^{2A-\phi} \phi_\pm)=c_+ e^{3A} * \lambda \, f\ , \label{eq:psp37f}
	\end{align}
\end{subequations}
which can indeed be found in \cite[Eq.~(2.5)]{haack-lust-martucci-t} (see also \cite{smyth-vaula}).

Finally we look at (\ref{eq:++1}), (\ref{eq:++2}). The computation is again very similar to the one in section \ref{ssub:mink4repr}. For $M=m$, $N=n$, and for $M=m$, $N=\nu$, we find nothing new. For $M=\mu$, $N=\nu$, instead of (\ref{eq:*fphi}) we now find, in terms of the three-dimensional Chevalley--Mukai pairing,
\begin{equation}\label{eq:fphi37}
	(f,\phi_\mp)=0\ .
\end{equation}
Unlike what happened in four dimensions, however, this equation cannot be derived from the pure spinor equations. It is possible to rewrite it as $(*\lambda f, \phi_\pm)=0$; but, if we try to derive this from (\ref{eq:psp37f}), we now get a term $(d \phi_\pm,\phi_\pm)$, about which we cannot in general say anything. Indeed (\ref{eq:fphi37}) was listed in \cite[Eq.~(2.6)]{haack-lust-martucci-t} as a separate algebraic constraint.

In conclusion, also for this class of solutions our system (\ref{eq:susy10}) reproduces the conditions for supersymmetry found in previous work --- in this case \cite{haack-lust-martucci-t,smyth-vaula}. The differential, pure-spinor-like equations are again all reproduced by (\ref{eq:psp10}); equations (\ref{eq:++1}), (\ref{eq:++2}) give an algebraic constraint on the flux that was also found in \cite{haack-lust-martucci-t,smyth-vaula}. We can regard~(\ref{eq:++1}),~(\ref{eq:++2}) as a generalization of that constraint.

\section*{Acknowledgments}
I would like to thank L.~Martucci, R.~Minasian, M.~Petrini and A.~Zaffaroni for interesting discussions. I am supported in part by INFN.



\appendix

\section{Bispinors} 
\label{sec:bispinors}

In this section we will collect a few facts about gamma matrices that we need in the main text and in appendix \ref{app:suff}. There is very little of substance here; we are mostly going to take care of a few annoying but unavoidable signs. 

\subsection{The $\lambda$ operator} 
\label{sub:lambda}
The annoying sign par excellence is the operator $\lambda$, already encountered in the main text:
\begin{equation}
	\lambda(C_k)\equiv (-)^{\lfloor \frac k2 \rfloor} C_k\ ,
\end{equation}
where the floor function $\lfloor \cdot \rfloor$ denotes the integer part. In this subsection, we are going to focus on $d=10$ in Lorentzian signature, with real gamma matrices as in section \ref{sub:eps}. 

$\lambda$ is related to transposition: from (\ref{eq:gammat}) we see that
\begin{equation}
	C_k^t = (-)^{k-1} \gamma^0 \lambda(C_k) \gamma^0 \ .
\end{equation} 
In particular we have
\begin{equation}\label{eq:e2e1}
	(-)^{{\rm deg}(\Phi)}\lambda(\Phi)= - \gamma^0 \Phi^t \gamma^0 = 
	- \gamma^0 (\gamma^0)^t \epsilon_2 \epsilon_1^t \gamma^0= - \epsilon_2 \overline{\epsilon_1}\ .
\end{equation}
Moreover, since $\Phi$ is even (odd) when $\epsilon_2$ is odd (even), we get
\begin{equation}\label{eq:glP}
	\gamma \lambda(\Phi) = -(-)^{{\rm deg}(\Phi)} \lambda(\Phi) \ .
\end{equation}
$\lambda$ does not commute with wedges and contractions: since $(-)^{\lfloor \frac {k+1}2 \rfloor}= (-)^k (-)^{\lfloor \frac k2 \rfloor}$, we have
\begin{equation}
	\lambda(dx^M \wedge C_k)= (-)^k dx^M \wedge \lambda(C_k)\ ,\qquad
	\lambda(\iota_M C_k)= (-)^{k+1} \iota_M C_k \ ,
\end{equation}
from which, remembering (\ref{eq:gwedge}), 
\begin{equation}\label{eq:lg}
	\lambda( \gamma^M C ) = \lambda(C) \gamma^M \ .
\end{equation}

\subsection{Hodge star} 
\label{sub:hodge}

We will now consider gamma matrices in any dimension. We will first consider the case of Euclidean signature. The chiral operator can be written as
\begin{equation}
	\gamma= c\, \gamma^1 \ldots \gamma^{d}\ ,
\end{equation}
where $c$ is a constant such that
\begin{equation}
	c^2= (-)^{\lfloor \frac d2 \rfloor}\ .
\end{equation}
(When $d$ is odd, we will take $\gamma=1$.) $\gamma$ is related to the Hodge star by
\begin{equation}\label{eq:stard}
	\gamma\, C_k = c * \lambda \,C_k \ ,\qquad
	C_k \gamma = c \,(-)^{\lfloor \frac d2 \rfloor} \lambda * C_k \ .
\end{equation}

In Lorentzian signature, there is an extra minus sign:
\begin{equation}
	\gamma= c\, \gamma^0 \ldots \gamma^{d-1}\ ,\qquad
	c^2=-(-)^{\lfloor \frac d2 \rfloor}\ .
\end{equation}
With this definition, (\ref{eq:stard}) still holds.

We also need the relation between the Chevalley--Mukai pairing (\ref{eq:mukai}) and the trace of bispinors: 
\begin{equation}\label{eq:mukaiTr}
	\frac1{2^{\lfloor \frac d2 \rfloor}}{\rm Tr}(*A \,B) = (-)^{{\rm deg}(A)} (A,B)\ .
\end{equation}

Finally, there are two formulas that we will use several times in the main text. One is the Fierz identity. For even dimension, it consists in expanding any bispinor $C$ on the basis $\{ \gamma^{M_1\ldots M_k}\}_{k=0}^d$:
\begin{equation}\label{eq:fierzd}
	C= \sum_{k=0}^d \frac1{2^{\frac d2} k!} {\rm Tr}(C \gamma_{M_k\ldots M_1}) \gamma^{M_1\ldots M_k}
	\ \ \Rightarrow \ \ \zeta_1 \otimes \overline{\zeta_2} =\sum_{k=0}^d \frac1{2^{\frac d2} k!} (\overline{\zeta_2}\gamma_{M_k\ldots M_1}\zeta_1) \gamma^{M_1\ldots M_k}\ .
\end{equation} 
For odd dimension, the matrices $\{ \gamma^{M_1\ldots M_k}\}_{k=0}^d$ are twice as many as the dimension of the space of bispinor, $2^{\lfloor \frac d2 \rfloor}\times 2^{\lfloor \frac d2 \rfloor}$. So there are in fact two bases: $\{ \gamma^{M_1\ldots M_k}\}_{k=0\, (k\ {\rm even})}^d$ and $\{ \gamma^{M_1\ldots M_k}\}_{k=0\,(k\ {\rm odd})}^d$. So we have two possibilities:
\begin{equation}\label{eq:fierzodd}
	C= \sum_{\substack{k=0\\ k\ {\rm even/odd}}}^d \frac1{2^{\lfloor \frac d2 \rfloor} k!} {\rm Tr}(C \gamma_{M_k\ldots M_1}) \gamma^{M_1\ldots M_k}
	\ \ \Rightarrow \ \ \zeta_1 \otimes \overline{\zeta_2} = \sum_{\substack{k=0\\ k\ {\rm even/odd}}}^d\frac1{2^{\lfloor \frac d2 \rfloor} k!} (\overline{\zeta_2}\gamma_{M_k\ldots M_1}\zeta_1) \gamma^{M_1\ldots M_k}\ .
\end{equation}
So when the dimension is odd the Clifford map (\ref{eq:cliffordmap}) is not injective: an odd form $C_-$ and an even form $C_+$ can correspond to the same bispinor, $\sla C_- = \sla C_+$. In fact, this happens when $C_+$ and $C_-$ are related by $c*\lambda$, as one can see from (\ref{eq:stard}) (recalling that, when $d$ is odd, $\gamma=1$).

The last formula we need in the main text is valid in any number of dimensions:
\begin{equation}\label{eq:gCgd}
	\gamma^M C_k \gamma_M = (-)^k (d-2k) \, C_k\ .
\end{equation}


\section{Equivalence of (\ref{eq:susy10}) to supersymmetry} 
\label{app:suff}

We will describe here the derivation of (\ref{eq:susy10}), and show that it is equivalent to the conditions for unbroken supersymmetry. We will work in IIB; the computations for IIA are very similar.

\subsection{Deriving (\ref{eq:psp10})} 
\label{sub:psp10}

The derivation of (\ref{eq:psp10}) is similar to the derivation of the pure spinor equations in~\cite[App.~A]{gmpt3}. A notable difference is that one cannot use the ``spacetime gravitino'' variation $\delta \psi_\mu$, since there is no distinction between spacetime and internal indices now. As we will see, we can proceed anyway, thanks to the properties of ten-dimensional spinors that we reviewed in section \ref{sub:eps}.

For supersymmetry to be unbroken, we want to set the supersymmetry variations to zero:
\begin{subequations}\label{eq:susy}
	\begin{align}
		&\left(D_M - \frac14 H_M\right) \epsilon_1 + \frac{e^\phi}{16} F \gamma_M \epsilon_2 = 0 
		\ ,&\qquad \left(D - \frac14 H - \del \phi\right) \epsilon_1 =0	
		\label{eq:susy1}\\
		&\left(D_M + \frac14 H_M\right) \epsilon_2 - \frac{e^\phi}{16} \lambda(F) \gamma_M \epsilon_1 = 0 
		\ ,&\qquad \left(D + \frac14 H - \del \phi\right) \epsilon_2=0 \ ,\label{eq:susy2}
	\end{align}
\end{subequations}
where, as usual, $D\equiv \gamma^M D_M $ is the Dirac operator, $H\equiv \sla H= \frac16 H_{MNP}\gamma^{MNP}$, $H_M=\frac12 H_{MNP} \gamma^{NP}$.

We start by computing $(d_H \Phi - d \phi \wedge \Phi)$. For $d \Phi= dx^M\wedge\del_M \Phi $ and $d\phi \wedge= \del_M \phi\, dx^M\wedge$, we can just derive $2dx_M\wedge= \stackrel \to \gamma_M + \stackrel \leftarrow \gamma_M (-)^{\rm deg}$ from (\ref{eq:garrow}). As for $H\wedge$, it can be obtained by applying the same expression to each of the three wedges in $dx^M\wedge dx^N \wedge dx^P\wedge $:
\begin{equation}
	\begin{split}
		H\wedge &= \frac1{8\cdot 6}H^{MNP}\left(\stackrel\to \gamma_{MNP} + \stackrel \leftarrow\gamma_{MNP}(-)^{\rm deg} + 3 \stackrel \to \gamma_M \stackrel \leftarrow \gamma_{NP}+3 \stackrel \to \gamma_{NP} \stackrel \leftarrow \gamma_M (-)^{\rm deg}\right)=\\
		&=\frac18 \left(\stackrel \to H + \stackrel \leftarrow H (-)^{\rm deg} + \stackrel \to \gamma^M \stackrel \leftarrow H_M + \stackrel \leftarrow \gamma^M \stackrel \to H_M (-)^{\rm deg} \right)\ .
	\end{split}
\end{equation}
So we can now write
\begin{equation}\label{eq:psp10dim}
	\begin{split}
		2\left(d_H \Phi - d \phi \wedge \Phi\right) &=[\gamma^M,\del_M \Phi - \del_M \phi \wedge \Phi] -2 H\wedge \Phi\\
		&=\left(D \epsilon_1 -\frac14 H \epsilon_1- \del \phi \epsilon_1 \right) \overline{\epsilon_2} 
		+ \gamma^M \epsilon_1\left(D_M \overline{\epsilon_2} -\frac14 \overline{\epsilon_2} H_M\right)\\
		&-\left(D_M \epsilon_1 - \frac14 H_M \epsilon_1\right) \overline{\epsilon_2} \gamma^M - \epsilon_1 \left( \overline{D\epsilon_2} -\frac14 \overline{\epsilon_2} H-\overline{\epsilon_2}\del \phi\right)\ .
	\end{split}
\end{equation}
Using (\ref{eq:susy}), this becomes
\begin{equation}\label{eq:psp10dim2}
	 \frac{e^\phi}{16} \left(\gamma^M \epsilon_1 \overline{\epsilon_1}\, \gamma_M F + F\, \gamma_M \epsilon_2\overline{\epsilon_2}\gamma^M \right)\ . 
\end{equation}
We can now use the property (\ref{eq:geeg}). Crucially, the right hand side of that relation only involves vectors and one-forms, so that we can massage (\ref{eq:psp10dim2}) without needing any extra equation (such as the external gravitino variation in \cite[App.~A]{gmpt3}): 
\begin{equation}\label{eq:psp10dim3}
	{\rm (\ref{eq:psp10dim2})}= -\frac12 e^\phi \left(K_1 (1+\gamma) F + F (1-\gamma)K_2\right)=-2 e^\phi \left(\tilde K\wedge + \iota_K\right) F\ .
\end{equation}
We have used that
\begin{equation}\label{eq:gF}
	\gamma F = F\ 
\end{equation}
(which follows from self-duality, (\ref{eq:F*F}), and from (\ref{eq:star})), and the definition of $K$ and $\tilde K$ in~(\ref{eq:KKt}). Comparing now (\ref{eq:psp10dim}) with (\ref{eq:psp10dim3}), we obtain (\ref{eq:psp10}).

As for (\ref{eq:LK}), those two equations have been already derived in \cite{hackettjones-smith,figueroaofarrill-hackettjones-moutsopoulos,koerber-martucci-ads}, so we will not give their derivation here.


\subsection{Spinor basis} 
\label{sub:completing}

In section \ref{ssub:structure}, we introduced a basis for ten-dimensional gamma matrices. As we noted there, the spinors 
\begin{equation}\label{eq:+basis}
	\{\gamma^{AB}\epsilon\}
\end{equation}
span the whole space of spinors with the same chirality as $\epsilon$. However, we will also need later a basis for the space of spinors with the opposite chirality. For ease of discussion, we will pick $\epsilon$ to be of chirality $+$, as we did in section \ref{ssub:structure}. So we need a basis for the space $\Sigma_-$ of spinors of negative chirality.

Obviously, $\gamma^M \epsilon$ cannot be enough, since there are only 9 of them (recall $\gamma^+ \epsilon=0$, from~(\ref{eq:g+e})), and the space $\Sigma_-$ has dimension 16. The next natural possibility is to use spinors of the form $\gamma^{MNP}\epsilon$:
\begin{equation}\label{eq:oppchir}
	\gamma^{\alpha \beta \gamma} \epsilon= | \uparrow \ \rangle \otimes \gamma_{(8)}^{\alpha \beta \gamma} \eta_+ \ ,\qquad
	\gamma^{- \alpha \beta} \epsilon = | \downarrow \ \rangle \otimes \gamma_{(8)}^{\alpha \beta}\eta_+ \ ,\qquad
	\gamma^{+- \alpha } \epsilon = 2\,| \uparrow \ \rangle \otimes \gamma^\alpha_{(8)} \eta_+ \ .
\end{equation}
However, in eight dimensions we have
\begin{equation}
	\gamma^{\alpha \beta \gamma} \eta_+ = \Psi^{\alpha \beta \gamma \delta} \gamma_\delta \eta_+ \ ,
\end{equation}
where $\Psi^{\alpha \beta \gamma \delta}$ is the ${\rm Spin}(7)$ four-form in eight dimensions. So the $\gamma^{\alpha \beta \gamma} \epsilon$ are in fact dependent on the $\gamma^{+-\alpha}\epsilon$, of which there are 8. We also know that there are only 7 non-vanishing $\gamma^{\alpha \beta}\eta_+$; so we are left with only $8+7=15$ spinors. The one which we are missing is 
\begin{equation}
	\gamma_+ \epsilon = | \downarrow \ \rangle \otimes \eta_+ \ .
\end{equation}

So neither $\{\gamma_M \epsilon\}$ nor $\{ \gamma_{MNP}\, \epsilon\}$ give a complete basis for $\Sigma_-$. One possibility would be to use them both, as a redundant basis. 

Another possibility, which we will adopt, is to pick a particular spinor with negative chirality, and act on this with the $\gamma^{MN}$. Given our discussion in section \ref{sub:metric}, a natural choice for this spinor is $\gamma_+ \epsilon$. Hence our basis for spinors with chirality opposite to $\epsilon$ is
\begin{equation}\label{eq:-basis}
	\{\gamma^{AB}\gamma_+ \epsilon\} \ .
\end{equation}


\subsection{Original supersymmetry equations in terms of intrinsic torsion} 
\label{sub:original}

We will now count the independent components of the supersymmetry equations, so that we can compare them in section \ref{sub:psp10QT} with (\ref{eq:psp10}) and (\ref{eq:LK}). 

We can now use the bases (\ref{eq:+basis}) and (\ref{eq:-basis}) for both $\epsilon_1$ and $\epsilon_2$. Just like in section \ref{sub:epseps}, we need to take care to distinguish the index $+$ relative to $\epsilon_1$ from the index $+$ relative to $\epsilon_2$, which we will by adding an index ${}_1$ or ${}_2$; and likewise for the indices $-$ and $\alpha$. So we will have indices $+_1, -_1, \alpha_1$, and $+_2, -_2, \alpha_2$.  

We can now define
\begin{subequations}\label{eq:QT}
	\begin{align}
		&\left(D_M - \frac14 H_M\right) \epsilon_1 = Q^1_{MNP} \gamma^{NP}\epsilon_1 \ ,&\left(D - \frac14 H - \del \phi\right) \epsilon_1 = T^1_{MN}\gamma^{MN} \gamma_{+_1} \epsilon_1
		\label{eq:QT1}\\
		&\left(D_M + \frac14 H_M\right) \epsilon_2 = Q^2_{MNP} \gamma^{NP}\epsilon_2 \ , &\left(D + \frac14 H - \del \phi\right) \epsilon_2 = T^2_{MN}\gamma^{MN} \gamma_{+_2} \epsilon_2 \ .
		\label{eq:QT2}
	\end{align}
\end{subequations}
There is no assumption here: the left hand sides are spinors that can be expanded on our basis, and the $Q$'s and $T$'s are the coefficients of this expansion. They are the ten-dimensional analogue of the coefficients introduced in \cite[App.~A.4]{gmpt3}. They can be thought of as parameterizing the intrinsic torsion of the generalized ${\rm ISpin}(7)$ structure defined by $\Phi$. Notice that some of their components do not multiply anything, and can be assumed to vanish:
\begin{equation}
	Q^a_{M \alpha_a +_a}=0 \ ,\qquad T^a_{\alpha_a -_a}=0\ .
\end{equation}
For the same reason, we can assume the $Q$ to be antisymmetric in their last two indices, and the $T$ to be antisymmetric.

We also need a basis for forms. Our basis (\ref{eq:+basis}), (\ref{eq:-basis}) also produces for us a basis for the space of bispinors: 
\begin{equation}\label{eq:bispbasis}
	\gamma_{MN} \epsilon_1 \otimes \overline{\epsilon_2} \gamma_{PQ}\ ,\qquad
	\gamma_{MN} \gamma_{+_1}\epsilon_1 \otimes \overline{\epsilon_2} \gamma_{+_2}\gamma_{PQ} \ ;\qquad
	 \gamma_{MN} \gamma_{+_1}\epsilon_1 \otimes \overline{\epsilon_2} \gamma_{PQ}\ ,\qquad
	\gamma_{MN} \epsilon_1 \otimes \overline{\epsilon_2} \gamma_{+_2}\gamma_{PQ}\ .
\end{equation}
In IIB, which is our focus in this appendix, the first two sets of generators are odd forms, and the second two are even; in IIA, the opposite would be true. Via the Clifford map~(\ref{eq:cliffordmap}), the basis (\ref{eq:bispbasis}) can also be used as a basis for forms. In this section, we will use it to expand $F$. In IIB, $F$ is an odd form; moreover, it is self-dual, so that we have (\ref{eq:gF}): $\gamma F = F$. This tells us that it will be a linear combination of the first set of generators in (\ref{eq:bispbasis}):
\begin{equation}\label{eq:FR}
	F=  R_{MNPQ} \gamma^{MN} \epsilon_1 \otimes \overline{\epsilon_2} \gamma^{PQ}\ .
\end{equation}
It also follows that $\lambda(F)= R_{MNPQ} \gamma^{QP} \epsilon_2 \otimes \overline{\epsilon_1} \gamma^{NM}$. 

Expanding (\ref{eq:susy}) in terms of the coefficients we just introduced in (\ref{eq:QT}), we get 
\begin{equation}\label{eq:susyQT}
	\begin{split}
		&Q^1_{MNP}=4 e^\phi R_{NPM-_2} \ ,\qquad  T^1_{MN}=0\ , \\
		&Q^2_{MNP}=4 e^\phi R_{-_1MNP} \ ,\qquad  T^2_{MN}=0\ . \\
	\end{split}
\end{equation}


\subsection{(\ref{eq:psp10}), (\ref{eq:LK}) in terms of intrinsic torsion} 
\label{sub:psp10QT}

We will now rewrite the equations in (\ref{eq:susy10}) in the language of the intrinsic torsions $Q$ and $T$, so as to compare them with (\ref{eq:susyQT}).

We start from (\ref{eq:psp10}). Using (\ref{eq:psp10dim}) and (\ref{eq:QT}), we can write
\begin{equation}\label{eq:psp10l}
\begin{split}
	2(d_H \Phi - d \phi \wedge \Phi)=T^1_{MN} \gamma^{MN}\gamma_{+_1} \Phi - Q^2_{MNP} \gamma^M \Phi \gamma^{NP} - Q^1_{MNP} \gamma^{NP} \Phi \gamma^M + T^2_{MN} \Phi \gamma_{+_2} \gamma^{MN}
\end{split}
\end{equation}
and
\begin{equation}\label{eq:psp10r}
	2(\tilde K \wedge + \iota_K ) F = (K_1 \cdot F+ F\cdot K_2)= 
	R_{MNPQ}(K_1^M \gamma^N \Phi \gamma^{PQ}+ \gamma^{MN} \Phi \gamma^P K_2^Q )\ ,
\end{equation}
where we have used $K_1^R \gamma_R \epsilon^1=0$, $\overline{\epsilon^2}\gamma_R K_2^R$ and $[\gamma_R, \gamma^{MN}]= 4 \delta_R^{[M} \gamma^{N]}$. So, comparing~(\ref{eq:psp10l}) and~(\ref{eq:psp10r}) with (\ref{eq:psp10}), we get\footnote{One also needs to use that, in the basis of section \ref{ssub:structure}, $\gamma^{\alpha+}\gamma_+ \epsilon=2 \gamma^\alpha \epsilon$, $\gamma^{+-}\gamma_+ \epsilon = \gamma^- \epsilon$, $\gamma^{+-}\epsilon=\epsilon$.}
\begin{subequations}\label{eq:pspQT}
	\begin{align}
			&Q^1_{MN \alpha_1}= 4 e^\phi R_{N \alpha_1 M-_2} \qquad (M \neq +_2)\ , \qquad \qquad T^1_{\alpha_1 \beta_1}=0\ ,\label{eq:pspQT1}\\
			&Q^2_{MN \alpha_2}= 4 e^\phi R_{-_1 M N \alpha_2} \qquad (M \neq +_1)\ , \qquad \qquad T^2_{\alpha_2 \beta_2}=0\ ,\label{eq:pspQT2}\\
			&Q^1_{\alpha_2 +_1 -_1} + T^2_{\alpha_2 +_2} = 4 e^\phi R_{+_1 -_1 \alpha_2 -_2}\ , \qquad\qquad 
			T^2_{+_2 -_2 } =-2 Q^1_{-_2 +_1 -_1} \ ,\label{eq:pspQT3}\\
			&Q^2_{\alpha_1 +_2 -_2} + T^1_{\alpha_1 +_1} = 4 e^\phi R_{-_1 \alpha_1 +_2 -_2}\ , \qquad\qquad 
			T^1_{+_1 -_1} = - 2 Q^2_{-_1 +_2 -_2} \label{eq:pspQT4}\ .		
	\end{align}
\end{subequations}
All of these equations are implies by (\ref{eq:susyQT}), as they should. But the converse is not true: we see, for example, that the components
\begin{equation}
	Q^1_{+_2 NP}\ ,\qquad Q^2_{+_1 NP}\ 
\end{equation}
never appear anywhere in (\ref{eq:pspQT}). This is because they would multiply $\gamma^{NP}\epsilon_1 \overline{\epsilon_2}\gamma^{+_2}$ and $\gamma^{+_1} \epsilon_1 \overline{ \epsilon_2} \gamma^{NP}$, which vanish because $\gamma^+ \epsilon=0$. 

We can now try to add more equations to (\ref{eq:psp10}). An obvious choice is (\ref{eq:LK}), which already appeared in \cite{hackettjones-smith,figueroaofarrill-hackettjones-moutsopoulos,koerber-martucci-ads}. In terms of the coefficients $Q$, the first in (\ref{eq:LK}) says $Q^1_{(MN)-_1}+Q^2_{(MN)-_2}=0$, while the second says $Q^1_{[MN]-_1}-Q^2_{[MN]-_2}=0$. So together they give
\begin{equation}\label{eq:LKQ}
	Q^1_{MN-_1}=-Q^2_{NM-_2}\ .
\end{equation}
Unfortunately, not even this helps us recover the supersymmetry conditions (\ref{eq:susyQT}): for example, the components $Q^1_{+_2 \alpha_1 \beta_1}$ and $Q^2_{+_1 \alpha_2 \beta_2}$ do not appear in either (\ref{eq:pspQT}) or (\ref{eq:LKQ}).

We note in passing that already (\ref{eq:pspQT}) and (\ref{eq:LKQ}) are enough to show that $L_K \phi=0$, as claimed in section \ref{sub:symm}. By simply using the definition of $T$ in (\ref{eq:QT}), we see that
\begin{equation}\label{eq:d+k1k2}
	e^{2 \phi} d^\dagger (e^{-2 \phi} K_i)= -4 T^i_{+_i -_i}\ ,\qquad i=1,2\ .
\end{equation}
Now, we can use (\ref{eq:LKQ}) to write $Q^1_{-_2 +_1 -_1}= - Q^2_{+_1 -_2 -_2}$, which is zero because the two last indices of $Q$ are antisymmetric by definition. So (\ref{eq:pspQT}) implies that $T^2_{+_2 -_2}=0$; we can argue in a similar way that $T^1_{+_1 -_1}=0$. Summing now (\ref{eq:d+k1k2}) with $i=1$ and $i=2$, and using the fact that $K$ is Killing (from (\ref{eq:LK})), we get
\begin{equation}\label{eq:LKphi}
	L_K \phi = 0 \ ,
\end{equation}
as claimed.

\subsection{The missing equations: (\ref{eq:++1}), (\ref{eq:++2})} 
\label{sub:missing}

In section \ref{sub:psp10QT}, we have expressed (\ref{eq:psp10}) and (\ref{eq:LK}) in terms of the ``intrinsic torsions'' $Q$, $T$ introduced in (\ref{eq:QT}), and unfortunately we have concluded that they are implied by the supersymmetry conditions (\ref{eq:susyQT}), but not equivalent to them. We will now find some differential equations that, once expressed in terms of (\ref{eq:QT}), will provide the missing components of (\ref{eq:susyQT}). 

The first thing to remark is that we could have expected a priori that (\ref{eq:psp10}) and~(\ref{eq:LK}) would not be equivalent to supersymmetry. In section \ref{sec:gen}, we found that $\Phi$ does not contain by itself enough data to determine a metric and $B$ field; we have to also provide the elements $\stackrel \to \gamma_{+_1}$ and $\stackrel \leftarrow \gamma_{+_2}$ of $T \oplus T^*$. So we expect the missing equations to contain them. 

Unfortunately, $\stackrel \to \gamma_{+_1}$ and $\stackrel \leftarrow \gamma_{+_2}$ are not bilinears of $\epsilon_1$ and $\epsilon_2$, and so there is no straightforward procedure to compute their derivatives. However, one can proceed indirectly; let us focus on $\stackrel \to \gamma_{+_1}$. First of all we notice that
\begin{equation}
	\overline{\epsilon_1} \gamma_{+_1} \gamma^M \gamma_{+_1} \epsilon_1 = 32 e_{+_1}^M \ .
\end{equation}
This allows us to consider $e_{+_1}$ as a bilinear of $\gamma_{+_1}\epsilon_1$. We need to compute, however, the covariant derivatives of this spinor. For our purposes, it will be enough to know that
\begin{equation}
	\left\{ \left( D - \frac14 H - \del \phi\right), v_M \gamma^M\right\} = 2 v^N \left(D_M - \frac14 H_M \right) + \sla dv + e^{2 \phi} d^\dagger (e^{-2 \phi} v)\ .
\end{equation} 
This implies
\begin{equation}\label{eq:Dg+e}
	\left( D - \frac 14 H - \del \phi\right) \gamma_{+_1}\epsilon_1 =  \left(-\gamma_{+_1} T^1_{MN} \gamma^{MN} \gamma_{+_1} + 2 Q^1_{+_1 NP} \gamma^{NP} + d e_{+_1}\cdot + e^{2 \phi } d^\dagger (e^{-2 \phi} e_{+_1})\right) \epsilon_1\ .
\end{equation}
We can now compute
\begin{equation}
\begin{split}
	32 e^{2 \phi } d^\dagger (e^{-2 \phi} e_{+_1})&= \overline{\epsilon_1} \gamma_{+_1}(D-\del \phi)(\gamma_{+_1} \epsilon_1)+ 
	\left( D_M( \overline{\epsilon_1}\gamma_{+_1})\gamma^M - \overline{\epsilon_1}\gamma_{+_1} \del \phi\right) \gamma_{+_1} \epsilon_1\\
	&=\overline{\epsilon_1}\Big[2Q^1_{+_1NP}[\gamma_{+_1},\gamma^{NP}]+[e_{+_1},de_{+_1}] +2 e^{2 \phi } d^\dagger (e^{-2 \phi} e_{+_1}) \gamma_{+_1}\Big] \epsilon_1\\
	&=32\Big[8 Q^1_{+_1 +_1 -_1}+ 2 \iota_{K_1} \iota_{e_{+_1}} de_{+_1}+ e^{2 \phi } d^\dagger (e^{-2 \phi} e_{+_1}) )\Big]\ .  
\end{split}
\end{equation}
So we get 
\begin{equation}
	Q^1_{+_1 +_1 -_1}= -\frac14 \iota_{K_1} \iota_{e_{+_1}} de_{+_1}\ ;
\end{equation}
and, going back to (\ref{eq:Dg+e}):
\begin{equation}\label{eq:eg+Dg+e}
	(\overline{\epsilon_1} \gamma_{+_1})\left(D- \frac14 H - \del \phi\right) (\gamma_{+_1}\epsilon_1)= 16 e^{2 \phi } d^\dagger (e^{-2 \phi} e_{+_1})\ .
\end{equation}

We are now ready to derive (\ref{eq:++2}). We can use the same strategy as in (\ref{eq:psp10dim}) to compute $(d_H - d \phi \wedge)$; with some manipulations we get
\begin{equation}\label{eq:++2der}
	\begin{split}
	&\overline{\epsilon_1}\gamma_{+_1}\Big[\{d_H - d \phi\wedge, \stackrel \to \gamma_{+_1}\} \epsilon_1 \overline{\epsilon_2}\Big]\gamma^{MN} \gamma_{+_2}  \epsilon_2 \\ 
	&=\overline{\epsilon_1}\gamma_{+_1}
	\left[\left(\left\{D-\frac14 H - \del \phi , \gamma_{+_1}\right\} \epsilon_1 \right)\overline{\epsilon_2}+
	\{ \gamma^P, \gamma_{+_1} \}\epsilon_1 \left(D_P\overline{ \epsilon_2}-\frac14 \overline{\epsilon_2} H_P\right) 
	\right]\gamma^{MN} \gamma_{+_2} \epsilon_2\\
	&=16 e^{2 \phi } d^\dagger (e^{-2 \phi} e_{+_1}) \cdot 64 K_2^{[M} e_{+_2}^{N]} + 32 \left(D_{+_2}\overline{\epsilon_2}-\frac14 \overline{\epsilon_2} H_{+_2}\right) \gamma^{MN} \gamma_{+_2} \epsilon_2	\ ;
	\end{split}
\end{equation}
we have used (\ref{eq:eg+Dg+e}) and the normalization $\overline{\epsilon_1} \gamma_{+_1} \epsilon_1=16$ (which follows from our conventions in section \ref{ssub:structure}). 

We now want to reexpress the last line of (\ref{eq:++2der}) in terms of the pairing (\ref{eq:mukai}). In general, for any $C$, using (\ref{eq:e2e1}) we have: 
\begin{equation}\label{eq:pre-egCge}
	\overline{\epsilon_1} \gamma_{+_1} C \gamma_{+_2} \epsilon_2 =
	 -(-)^{{\rm deg}(\Phi)}{\rm Tr}(\gamma_{+_2} \lambda(\Phi) \gamma_{+_1}C )\ .
\end{equation}
We also have, using (\ref{eq:star}), (\ref{eq:lg}) and (\ref{eq:glP}):
\begin{equation}\label{eq:*gPg}
	*(\gamma_{+_1} \Phi \gamma_{+_2})= \gamma \lambda(\gamma_{+_1} \Phi \gamma_{+_2})=\gamma \gamma_{+_2} \lambda(\Phi) \gamma_{+_1}=
	(-)^{{\rm deg}(\Phi)}\gamma_{+_2}\lambda(\Phi) \gamma_{+_1}\ .
\end{equation}
Putting (\ref{eq:pre-egCge}) together with (\ref{eq:*gPg}), and using the relation (\ref{eq:mukaiTr}) between the Chevalley--Mukai pairing and the spinorial trace, we obtain
\begin{equation}\label{eq:egCge}
	\overline{\epsilon_1} \gamma_{+_1} C \,\gamma_{+_2} \epsilon_2 = 
	-32 (-)^{{\rm deg}(\Phi)} (\gamma_{+_1} \Phi\, \gamma_{+_2}, C)  \ .
\end{equation}
In particular, for $C= \Phi \gamma^{MN}$:
\begin{equation}\label{eq:gPgPg}
	(\gamma_{+_1}\Phi\, \gamma_{+_2}, \Phi \gamma^{MN}) = -32 K_2^{[M} e_{+_2}^{N]}(-)^{{\rm deg}(\Phi)}\ .
\end{equation}

We can now use the conjugate of the first equation (the gravitino variation) in (\ref{eq:QT2}), to rewrite the second term in the last line of (\ref{eq:++2der}) in terms of $F$:
\begin{equation}\label{eq:secondterm}
	32 \left(D_{+_2}\overline{\epsilon_2}-\frac14 \overline{\epsilon_2}\, H_{+_2}\right) \gamma^{MN} \gamma_{+_2} \epsilon_2=
	2 e^\phi \overline{\epsilon_1} \gamma_{+_1} F \gamma^{MN} \gamma_{+_2}\epsilon_2 \ .
\end{equation}
 Using also (\ref{eq:egCge}) and (\ref{eq:gPgPg}), we obtain (\ref{eq:++2}).


\subsection{The missing equations and intrinsic torsion} 
\label{sub:missingQT}

Finally, we want to express (\ref{eq:++2}), which we have derived in section \ref{sub:missing},  in terms of the ``intrinsic torsions'' $Q$ and $T$ introduced in (\ref{eq:QT}). Once we do that, we will be able to check whether the system (\ref{eq:susy10}) really is equivalent to the supersymmetry conditions~(\ref{eq:susy}). 

We need to go back to the last line of (\ref{eq:++2der}). The crucial term is the second one, which we have already expressed in terms of $F$ in (\ref{eq:secondterm}). We now follow a different route, and we express it in terms of $Q$ using (\ref{eq:QT}):
\begin{equation}\label{eq:secondtermQ}
	-32 Q^2_{+_1 PQ} \overline{\epsilon_2} \gamma^{PQ} \gamma^{MN} \gamma_{+_2} \epsilon_2 \equiv -32 Q^2_{+_1 PQ} \tilde P_2^{PQMN} \ .
\end{equation}
The tensor $\tilde P$ is not quite a projector, but it is diagonal, if viewed as a matrix whose first index is the pair of indices $PQ$, and the second index is the pair of indices $MN$. Specifically, we find that the only non-zero entries are (up to antisymmetry in $PQ$ and $MN$ separately):
\begin{equation}\label{eq:Ptilde}
	\tilde P_{\alpha \beta}{}^{\gamma \delta}= P^\mathbf{7}_{\alpha \beta}{}^{\gamma \delta} \ ,\qquad
	\tilde P_{\alpha -}{}^{\gamma -}=\frac14 \delta_\alpha^\gamma
	\ ,\qquad
	\tilde P_{+-}{}^{+-}= \frac18\ ,
\end{equation}
where $P^\textbf{7}$ is the projector in the $\mathbf{7}$ of Spin(7). 
We have omitted the indices ${}_2$ (writing $\alpha$ rather than $\alpha_2$, and so on) to make the equation more readable, and because the same equation is true for the spinor $\epsilon_1$ (and in fact this is needed for the analysis of (\ref{eq:++1}), which we are not giving here). 

If we now decompose $F$ as (\ref{eq:FR}) in (\ref{eq:secondterm}), and we compare to (\ref{eq:secondtermQ}), we get
\begin{equation}
	(Q^2_{+_1 PQ} - 4 e^\phi R_{-_1 +_1 PQ}) \tilde P_2^{PQMN}=0 \ .
\end{equation}
But the only possible values for the indices $PQ\ $ in both $Q^2_{+_1 PQ}$ and $R_{-_1 +_1 PQ}$ are $\alpha_2 \beta_2$, $\alpha_2 -_2$ and $+_2 -_2$. So, in fact, applying (\ref{eq:Ptilde}) to $\tilde P_2$ (that is, reinstating the indices ${}_2$: $\alpha\to \alpha_2$ and so on) tells us that
\begin{equation}\label{eq:Q2+}
	Q^2_{+_1 PQ} = 4 e^\phi R_{-_1 +_1 PQ}
\end{equation}
These are precisely the components we were missing in (\ref{eq:pspQT}), as we observed there. A similar analysis for (\ref{eq:++1}) also gives us
\begin{equation}\label{eq:Q1+}
	Q^1_{+_2 PQ} = 4 e^\phi R_{PQ +_2 -_2}\ .
\end{equation}
Together with (\ref{eq:pspQT1}) and (\ref{eq:pspQT2}), these give us the supersymmetry conditions (\ref{eq:susyQT}) for $Q^1_{MN \alpha_1}$ and $Q^2_{MN \alpha_2}$. We still seem to face a problem: (\ref{eq:pspQT3}) and (\ref{eq:pspQT4}) do not quite tell us what 
\begin{equation}
	Q^1_{\alpha_2 +_1 -_1}\ ,\qquad Q^1_{-_2 +_1 -_1}\ ,\qquad Q^2_{\alpha_1 +_2 -_2}\ ,\qquad Q^2_{-_1 +_2 -_2}
\end{equation}
are, because of the contributions from $T$. However, (\ref{eq:LKQ}) relates these components to 
\begin{equation}
	-Q^2_{+_1 \alpha_2 -_2}\ ,\qquad -Q^2_{+_1 -_2 -_2}=0\ ,\qquad -Q^1_{+_2 \alpha_1 -_1}\ ,\qquad -Q^1_{+_2 -_1 -_1}=0\ .
\end{equation}
The two non-zero components, $Q^2_{+_1 \alpha_2 -_2}$ and $Q^1_{+_2 \alpha_1 -_1}$, have already been accounted for in~(\ref{eq:Q2+}) and~(\ref{eq:Q1+}). This allows us to derive all the missing equations in~(\ref{eq:susyQT}).

In conclusion, we have shown that the system (\ref{eq:susy10}) is equivalent to the conditions~(\ref{eq:susy}) for unbroken supersymmetry.


\providecommand{\href}[2]{#2}
\end{document}